\documentstyle[12pt,epsfig]{article}
\newcommand{\pslash}{\not{\!P}}
\newcommand{\pnslash}{\not{\!P_N}}
\newcommand{\snslash}{\not{\!S_N}}
\newcommand{\slslash}{\not{\!S_L}}

\newcommand{\cinvslash}{\not{\!C_{s's}}}

\newcommand{\be}{\begin{equation}}
\newcommand{\ee}{\end{equation}}
\newcommand{\ba}{\begin{eqnarray}}
\newcommand{\ea}{\end{eqnarray}}

\newcommand{\nsigma}{\mbox{\boldmath $\sigma$}}
\newcommand{\ngamma}{\mbox{\boldmath $\gamma$}}

\newcommand{\nzero}{\mbox{\boldmath $0$}}

\newcommand{\nl}{{\bf      l}}
\newcommand{\nn}{{\bf      n}}

\newcommand{\nk}{{\bf      k}}
\newcommand{\np}{{\bf      p}}       
\newcommand{\nq}{{\bf      q}}
\newcommand{\nr}{{\bf      r}}         
\newcommand{\ns}{{\bf      s}}

\newcommand{\nP}{{\bf      P}}

\begin{document}
\begin{titlepage}
\mbox{} 
\vspace*{2.5\fill} 
{\Large\bf 
\begin{center}
%
RPWIA analysis of dynamical relativistic contributions in $^{16}O(\vec{e},e'\vec{p})$ reactions
%
\end{center}
} 
\vspace{1\fill} 
\begin{center}
{\large 
M.C. Mart\'{\i}nez$^{1}$,
J.A. Caballero$^{1}$,
T.W. Donnelly$^{2}$
}
\end{center}
\begin{small}
\begin{center}
$^{1}${\sl 
Departamento de F\'\i sica At\'omica, Molecular y Nuclear \\ 
Universidad de Sevilla, Apdo. 1065, E-41080 Sevilla, SPAIN 
}\\[2mm]
$^{2}${\sl 
Center for Theoretical Physics, Laboratory for Nuclear Science 
and Department of Physics\\
Massachusetts Institute of Technology,
Cambridge, MA 02139, USA 
}
\end{center}
\end{small}

\kern 1. cm \hrule \kern 3mm 

\begin{small}
\noindent
{\bf Abstract} 
\vspace{3mm} 

Coincidence scattering of polarized electrons from nuclei with
polarization transfer to outgoing nucleons is studied within the context of relativistic
mean field theory. Effects introduced by the dynamical enhancement
of the lower components of the bound nucleon wave function are analyzed
for the polarized response functions and transferred polarization asymmetries
assuming the relativistic plane-wave impulse approximation (RPWIA). Results obtained
by projecting out the negative-energy components are compared with the
fully-relativistic calculation for proton knockout from $p_{1/2}$ and
$p_{3/2}$ shells in $^{16}$O for a variety of kinematic situations. The
crucial role played by the relativistic dynamics
in some spin-dependent observables is clearly manifested 
even for low/medium values of the missing momentum.
The degree to which knowledge about nucleon form factors can be extracted from analyses of this type of process is also discussed.

\kern 2mm 

\end{small}

\kern 2mm \hrule \kern 1cm
\noindent MIT/CTP\#3215
\end{titlepage}




\section{Introduction}


Over the years analyses of quasielastic coincidence electron
scattering reactions have provided important insight into single-particle properties of
nuclei, in particular, on the energies, momentum
distributions and spectroscopic factors of nucleons in nuclei.
Implicit in these analyses is the assumption that for quasielastic
kinematics the reaction mechanism underlying $(e,e'N)$ reactions can be
treated with confidence in the impulse approximation (IA), i.e., assuming
the virtual photon attaches to a single bound nucleon that absorbs the
whole momentum ($q$) and energy ($\omega$), subsequently being ejected and detected
(see \cite{Bof96,Kel96,Frul85} for details).

The simplest approach to studies of $(e,e'N)$ reactions invokes the
standard plane-wave impulse approximation (PWIA). There the $(e,e'N)$ 
differential cross section factorizes into a `single-nucleon'
cross section describing the electron-nucleon scattering process, and a
spectral function that gives the probability for finding a nucleon in the
target nucleus with selected values of energy and momentum compatible with
the kinematics of the process. This factorization property makes $(e,e'N)$
reactions so appealing for investigations of nuclear structure, since it implies that single-particle distributions can in principle be probed in great detail.

Unfortunately, the PWIA is an oversimplified description of $(e,e'N)$
processes. Firstly, distortion of both electron and nucleon wave functions
due to electromagnetic and strong interactions with target and residual
nuclei are necessary in order to perform detailed comparisons with experiment. Although these new
ingredients in general destroy the factorization result, the
interpretation of experimental data is still usually based on this property
by defining an effective
spectral function and an effective nucleon momentum distribution (or reduced
cross section) that in PWIA corresponds exactly to the single-particle
density in momentum space.

Secondly, the PWIA itself involves a non-relativistic truncation, and for years most theoretical work on $(e,e'N)$ has been
carried out on the basis of particular non-relativistic approximations~\cite{Bof96,Frul85}. For instance, this is the
case for the standard distorted-wave impulse approximation (DWIA) that is involved 
in performing comparisons with experiment which uses non-relativistic expressions 
for the nucleon current operator and
wave functions. There, the current operator is evaluated 
by using a direct Pauli reduction involving expansions in powers of
$p/M_N$, $q/M_N$ and $\omega/M_N$, where $p$ is the missing momentum and
$M_N$ the nucleon mass, and the bound and scattered wave functions correspond
to Schr\"odinger solutions with phenomenological non-relativistic
potentials. Although DWIA has been satisfactorily used to describe 
$(e,e'N)$ experiments performed during 1970 and 1980's~\cite{Bof96}, its validity for the
interpretation of experiments performed for the last decade where higher energies have become available is 
questionable. The values of the momenta and energies involved in these
processes are high enough to invalidate the non-relativistic
expansions assumed in DWIA and new relativistic analyses of
$(e,e'N)$ reactions become necessary.

A concerted effort has been made in recent years to incorporate
relativity in the description of coincidence $(e,e'N)$ 
processes~\cite{Pick87,Pick89,Hedayati,Kel97,Jin}.
Within the framework of the relativistic mean field
approach, nuclear responses and differential
cross sections for quasielastic coincidence electron scattering
have been investigated~\cite{Udi93,Udi95,Udi96,Udi99,Udi01}.
In the relativistic distorted-wave impulse approximation (RDWIA), bound and
scattering wave functions are described by solutions of the Dirac equation
with scalar and vector (S-V) potentials, and use is made of
the relativistic free-nucleon current operator. By comparing 
standard DWIA and RDWIA calculations,
relativistic contributions can then be cast into two general
categories, kinematical and dynamical relativistic effects. The former
are directly connected with the structure of the 
4-vector current operator, compared with the non-relativistic one that usually
involves $p/M_N$, $q/M_N$ and $\omega/M_N$
expansions. In some studies~\cite{Ama96,Ama98,Jesch}, new `relativized' 
current operators have been derived by making only expansions in 
$p/M_N$, but not in $q/M_N$ or $\omega/M_N$. These new expressions
retain important aspects of relativity not taken into account in the
traditional non-relativistic approximation and in so-doing these
should perhaps be called ``semi-relativistic'' approaches. 
The latter, dynamical relativistic effects, come from the
difference between the relativistic and non-relativistic 
nucleon (bound and ejected) wave functions involved.
Within these dynamical relativistic effects one may distinguish
effects associated with the Darwin term (that 
mainly affects the determination of spectroscopic factors at low 
missing momenta) and effects due to the dynamical enhancement of 
the lower components of the relativistic wave functions 
(which are expected to be especially relevant at high 
missing momenta, although they have proven to play an important role for
some particular observables even at low/medium $p$ values).
So far, fully-relativistic analyses of $(e,e'p)$ reactions have clearly 
improved the comparison with experimental data. In~\cite{Udi93,Udi95}
it was shown that the non-local Darwin term causes an enhanced absorption when
comparing RDWIA cross section with the DWIA one at moderate $p$ values,
thus predicting larger spectroscopic factors~\cite{Udi93,Udi01}. 
For larger missing momenta the lower components of the relativistic wave functions start to
play a more important role, enhancing the high momentum components of the 
nucleon wave functions. RDWIA calculations, compared with DWIA, produce
larger cross sections at $p>300$ MeV/c, also improving agreement with
experiment~\cite{Udi96}.

In parallel with such RDWIA calculations, we have also undertaken a more systematic
study of the effects of the dynamical enhancement of the lower components
within RPWIA, i.e., neglecting final-state interactions (FSI) between the
outgoing nucleon and the residual nucleus. Although a description
of FSI is necessary to analyze experimental data in detail, the RPWIA 
approach allows one to simplify the analysis
of the relativistic effects, disentangling them from other distortion effects.
Moreover, within RPWIA it is possible to get analytical expressions
that explicitly incorporate the contributions in the various observables
coming from the negative-energy components of the bound nucleon wave
function. This subject was throughly developed in~\cite{Cab98a}
(see also~\cite{Gardner94}) for
the case of unpolarized $A(e,e'N)B$ reactions. Nuclear response functions
and cross sections were evaluated and compared with
the standard PWIA-predictions.
The effects introduced by the presence of negative-energy components in the 
relativistic bound nucleon wave function were shown to be very important
for some observables even at low/moderate values of the missing momentum.
In particular, the interference
$TL$ and $TT$ responses were
shown to be the most sensitive observables to
dynamical effects of relativity affecting the lower components.
These results have been shown also to
persist in modeling where FSI are included. In
fact, data obtained on $R^{TL}$ and the left-right asymmetry $A_{TL}$ provide a strong indication of the
crucial role played by dynamical relativistic effects in $(e,e'p)$ reactions~\cite{Udi99,Udi01}.

The main aim of this paper is to analyze whether these dynamical relativistic effects 
may also have a significant impact on the polarization observables. Since spin and relativity go hand in hand, one may {\it a priori} consider
the relativistic approach to be better suited to describe nucleon polarization
observables. As is well known, when polarization
degrees of freedom are involved, a much richer variety of observables becomes
accesible. These contain in general interferences between various amplitudes and 
consequently a complete decomposition into the electromagnetic matrix elements can in
principle be achieved~\cite{RaDo89}. In this work we focus on the case of final-state
nucleon polarization measurements, i.e., $A(\vec{e},e'\vec{N})B$
processes, and we work within the context of the RPWIA.
The analysis of polarized observables including final-state interactions 
within RDWIA will be presented in a forthcoming publication. Thus, this work follows
closely the formalism presented in~\cite{Cab98a} for the unpolarized responses.

In recent years a number of experiments have been proposed or carried out 
to measure the polarization of the ejected nucleon in
$(\vec{e},e'\vec{p})$ reactions~\cite{Mal00,Die01}. Specifically, a concerted experimental
effort has been made with the aim of shedding some light on the issue of
the form factors of nucleons inside the nuclear medium. Indeed, recoil nucleon
polarization observables may be well-suited to provide valuable information on the nucleon form 
factors~\cite{Mal00}. Such studies present clear
advantages compared with the usual Rosenbluth separation method, as they do not require one to vary the electron beam energy and/or the scattering
angle, thus eliminating the systematic uncertainties that make it so difficult 
to extract the electric nucleon form factor $G_E$ at high $|Q^2|$ using the Rosenbluth method.
Furthermore, recoil polarization calculations for low/medium missing momenta 
have also proven to be relatively insensitive
to different ingredients in the description of the reaction mechanism, namely off-shell
ambiguities and optical potentials used to describe FSI~\cite{Kel97}.

The first measurements of recoil nucleon polarization were performed 
at Bates~\cite{Ede94,Mil98} and Mainz~\cite{Eyl95} analyzing 
\( ^{2}H(\vec{e},e'\vec{n}) \) and \( ^{2}H(\vec{e},e'\vec{p})n \) reactions. 
From these experiments a first estimation of 
the neutron and proton form factors was given at
several values of \( Q^{2} \).
Recently, high precision polarization transfer measurements on
complex nuclei have been presented by Malov 
et al. in $^{16}O(\vec{e},e'\vec{p})^{15}N$~\cite{Mal00},
and by Dieterich et al. in $^4He(\vec{e},e'\vec{p})^3H$~\cite{Die01}.
Although the general conclusions in these two last experiments are not free from
ambiguities due to experimental uncertainties, the authors in~\cite{Die01}
show that standard non-relativistic calculations are in clear disagreement with the 
experimental data. This result constitutes a strong indication of the necessity for
a fully-relativistic calculation in order to describe the spin transfer observables. 
Moreover, it also agrees with the general analysis presented in the case of induced
polarization measurements~\cite{Udi00,Woo98,JoSh99}, where the fully RDWIA calculation clearly provides a better description of the experimental data when compared with the standard non-relativistic
DWIA analysis. 

In this paper we present a systematic study within RPWIA of the new response functions that enter in the description of
$A(\vec{e},e'\vec{N})B$ processes. Following the arguments presented for the
unpolarized case in~\cite{Cab98a}, here we extend the analysis
to the polarized situation and for the polarized responses attempt to
identify clear signatures that arise from the dynamical relativistic effects coming from the
negative-energy projections (NEP) of the relativistic bound nucleon wave 
function. Moreover, the role played by the NEP on the
transferred polarization is also analyzed in detail. 
The work presented here is being undertaken in concert with studies of
related issues involving relativistic effects in coincidence electron
scattering with active polarization degrees of freedom. To attempt to
integrate all of the work into one manuscript would be far too
unwieldy and hence we have found what we hope are natural division
points to subdivide the presentations. In the light of this it is
important to understand how the various studies interact with one
another:
\begin{itemize}
\item In an accompanying paper~\cite{Cris1} the focus is placed on {\bf kinematical} relativistic
  effects for responses and polarization observables, however strictly
  within the context of the PWIA. Both off-shell and `on-shell'
  approaches are taken.  Basic essential formalism is also collected and presented in that paper.
\item In contrast in the present work we have emphasized the roles
  played by {\bf dynamical} relativistic effects which arise from
  non-trivial relativistic content in the nuclear wave functions. The
  present study has been limited to initial-state dynamical effects
  and thus the final state is still treated as a relativistic plane
  wave, the so-called RPWIA.
\item In work in progress effects in the final state are
  also being incorporated through {\bf relativistic FSI}, constituting the RDWIA. Upon
  completing this third phase of the overall study meaningful
  comparisons with measured observables can be attempted.
\end{itemize}

The paper is organized as follows: in Section~2 we
describe $A(\vec{e},e'\vec{N})B$ reactions within the context of the relativistic
plane-wave impulse approximation. Here a separation of the polarized response
functions and cross section into positive- and negative-energy 
projections is made. In Section~3 we present and discuss the results obtained for various
kinematic situations. The role of the negative-energy components and their influence on 
various choices of the current operator and/or gauge in the single-nucleon responses is
presented in Section~3.1. The total polarized hadronic responses and transferred polarization
asymmetries are discussed in Sections 3.2 and 3.3, respectively. Finally, in Section~4 we summarize
our main conclusions.


\section{Analysis of $A(\vec{e},e'\vec{N})B$ reactions in the Relativistic Plane-Wave Impulse
Approximation}


In the $(\vec{e},e'\vec{N})$ process a longitudinally polarized electron with 4-momentum 
\( K=(\varepsilon,\nk) \) is scattered through an angle $\theta_e$ to a 4-momentum
\( K'=(\varepsilon',\nk') \). The hadronic variables are denoted by
\( P_{A}^\mu=(M_{A},\nzero) \), \( P_{N}^\mu=(E_{N},\np_{N}) \) and 
\( P_{B}^\mu=(E_{B},\np_{B}) \) representing the 4-momenta of the target, outgoing nucleon and residual nucleus, respectively. The 4-momentum transfer is given by
\( Q^\mu=K^\mu-K'^\mu=(\omega ,\nq) \).
Within the impulse approach (IA), the virtual photon is absorbed by
a single nucleon in the nucleus whose 4-momentum is
\( P^\mu=P_N^\mu-Q^\mu=(E,\np) \). 
As usual, electrons are treated in the extreme relativistic limit (ERL), i.e.,
$\varepsilon=k$, $\varepsilon'=k'$.

The general formalism for coincidence electron scattering reactions with the electron beam polarized 
and with the outgoing nucleon's polarization measured has been presented in detail in previous
work~\cite{Bof96,Pick87,Pick89,RaDo89,Giusti89} -- in particular, as noted above, 
the present work has been
undertaken in concert with a focused study of kinematic relativistic effects in
$(\vec{e},e'\vec{N})$ reactions~\cite{Cris1} and the reader is directed to that
accompanying paper for detailed discussions of our conventions. 
Here we restrict our attention to the 
Relativistic Plane-Wave Impulse Approximation (RPWIA) where the analysis of 
$(\vec{e},e'\vec{N})$ reactions is further simplified. In particular, the
analyzing power $A$ and induced polarization $\nP$ (see~\cite{Pick87,Pick89} for details)
are zero in that case, and only the transferred polarization asymmetry $\nP'$ survives.
In terms of nuclear responses,
from the total of eighteen response functions that enter in
the general analysis of $A(\vec{e},e'\vec{N})B$ reactions~\cite{Pick87,Pick89} 
only nine survive within RPWIA. Four ($R^L_0$, $R^T_0$, $R^{TL}_0$ and $R^{TT}_0$) represent 
the unpolarized responses and
the five remaining ($R^{T'}_l$, $R^{T'}_s$, $R^{TL'}_l$, $R^{TL'}_s$ and $R^{TL'}_n$) depend explicitly
on the recoil nucleon polarization and only enter when the electron beam is also polarized. The 
polarized responses are referred to the coordinate system defined by the unit
vectors $(\nl,\ns,\nn)$ with\footnote{Here we use $l$, $n$ and $s$ to
  label the three axes, whereas in~\cite{RaDo89} different conventions were
  adopted: there unprimed labels were reserved for initial-state
  polarizations and primed labels ($l'$, $n'$, $s'$) were employed
  when final-state polarizations occur as here.}
$\nl$ (parallel to the momentum 
$\np_N$ of the outgoing nucleon), $\nn$ (perpendicular to the plane 
containing $\np_N$ and the transfer momentum $\nq$), and $\ns$ 
(determined by $\nn \times \nl$).

Although the plane-wave limit represents an oversimplified description of electron scattering
reactions, our goal in this paper is not so much to compare the calculation with experimental data, for which
FSI are surely necessary, but rather is to focus on the dynamical relativistic effects coming from the presence of negative-energy
projections of the bound nucleon wave function, and to investigate how this enhancement of the lower components
affects the recoil nucleon polarized
responses. In this sense, the simplification of invoking the
plane-wave limit may help us gain important insight when trying
to disentangle effects of distortion from effects of high $p$ components in the bound
nucleon wave function and current operator. Moreover, the RPWIA cross section and response
functions can be separated into contributions from positive- and
negative-energy projections of the bound nucleon wave function. These contributions
can be analyzed separately to yield a clear indication of how the factorization limit of
the standard PWIA analysis breaks down.

The RPWIA formalism applied to the analysis of the unpolarized
$(e,e'N)$ reaction was developed
in detail in~\cite{Cab98a}. Here we apply the same approach to the case of recoil nucleon
polarization measurements, and hence simply summarize the basic ingredients needed to describe the
new observables, referring the reader to~\cite{Cab98a} for further details.
We use the conventions of~\cite{BjDr64}.

The cross section is proportional to the contraction of the leptonic tensor
$\eta_{\mu\nu}$ with the hadronic tensor
$W^{\mu\nu}$ (see~\cite{Cris1}): $\eta_{\mu\nu}$ can be evaluated from the electron
current matrix elements (see for instance~\cite{RaDo89}) and $W^{\mu\nu}$ arises from bilinear
combinations of the nuclear current matrix elements.
Within RPWIA, 
requiring that the recoil nucleon polarization be measured, one can write for the latter
\be
W^{\mu\nu}= \frac{2}{2j+1}\sum_{m}
        \left[\overline{u}(\np_N,s_{N})\hat{J}^{\mu}\Psi_{\kappa}^{m}(\np)\right]^\ast
        \left[\overline{u}(\np_N,s_{N})\hat{J}^{\nu}\Psi_{\kappa}^{m}(\np)\right] \, ,
\label{eq13}
\ee
where \( \Psi_{\kappa}^{m}(\np) \) denotes the Fourier transform of the relativistic
bound nucleon wave function \( \Psi_{\kappa}^{m}(\nr) \) which is a solution
of the Dirac equation with S-V potentials. The indices $\kappa$ and $m$ denote the
Dirac wave function quantum
numbers (see Appendix A in~\cite{Cab98a} for details).
An interacting relativistic wave function always presents a coupling to the free negative-energy
Dirac spinors $v$. Letting $\overline{E}$ denote the on-shell energy, i.e.,
$\overline{E}=\sqrt{p^2+M_N^2}$, this implies that the free relation between upper ($u$) and lower ($d$) components,
\be 
\frac{\nsigma\cdot\np}{\overline{E}+M_N}\phi^u=\phi^d \, ,
\label{eq13a}
\ee
does not hold in general for a bound relativistic wave function.  This is what
differentiates the RPWIA from the standard PWIA analysis. According to~\cite{Cab98a} 
(see also Appendix A) the nucleon
current matrix elements can be split into two terms, one coming from the positive-energy projector
involving the spinors $u(\np ,s)$, and the other from the negative-energy projector involving
the spinors $v(\np ,s)$. At the level of the hadronic tensor the following decomposition
results:
\be
W^{\mu\nu}=W^{\mu\nu}_{P}+W_{N}^{\mu\nu}+W_{C}^{\mu\nu}\, ,
\label{eq14}
\ee
where $W^{\mu\nu}_P$ ($W^{\mu\nu}_N$) is the contribution from positive- (negative-) energy
projections only, while $W^{\mu\nu}_C$ is a crossed term containing products of both positive-
and negative-energy projections. Following the arguments developed originally
in~\cite{Cab98a} and presented also for the case of recoil nucleon polarization
in Appendix A, each of the contributions in Eq.~(\ref{eq14}) can be factorized into two
terms, a single-nucleon tensor and a momentum distribution component,
\ba
W_{P}^{\mu\nu}&=&N_{uu}(p)\cal {W}^{\mu\nu} \label{eq15} \\
W_{N}^{\mu\nu}&=&N_{vv}(p)\cal {Z}^{\mu\nu} \label{eq16} \\
W_{C}^{\mu\nu}&=&N_{uv}(p)\cal {N}^{\mu\nu} \label{eq17} \, .
\ea
Note, however, that the total tensor in Eq.~(\ref{eq14}) is a sum of three terms and does not
in general factorize. 
Explicit expressions for the momentum distribution components $N_{uu}(p)$, $N_{vv}(p)$ and $N_{uv}(p)$ 
in terms of the upper and lower components of the bound nucleon wave function are given in~\cite{Cab98a} and are also summarized in Appendix A. Let us recall that by imposing the
free relation in Eq.~(\ref{eq13a}), only the contribution $N_{uu}(p)$ survives, i.e., all of the terms
containing negative-energy projections become zero.  In this limit factorization is recovered. 

The single-nucleon tensors ${\cal W}^{\mu\nu}$, ${\cal Z}^{\mu\nu}$ and ${\cal N}^{\mu\nu}$ are
given by
\ba
{\cal W}^{\mu \nu }&=&\frac{1}{8M_{N}^{2}}Tr\left[ (\pslash +M_{N})
        \overline{J}^{\mu }(\pnslash +M_{N})(1+\gamma _{5}\snslash )J^{\nu }\right] \label{eq23} \\
{\cal Z}^{\mu \nu }&=&\frac{1}{8M_{N}^{2}}Tr\left[ (\pslash -M_{N})
        \overline{J}^{\mu }(\pnslash +M_{N})(1+\gamma _{5}\snslash )J^{\nu }\right]\label{eq24} \\
{\cal N}^{\mu \nu }&=&\frac{1}{8M_{N}^{2}}Tr\left[ \overline{J}^{\mu }
        (\pnslash +M_{N})(1+\gamma _{5}\snslash )J^{\nu }\gamma ^{0}\frac{\ngamma
        \cdot \np}{p}\pslash \right] \label{eq25}\, ,
\ea
where we use the notation $\overline{J}^\mu\equiv \gamma_0 J^{\mu +} \gamma_0$. 

The term ${\cal W}^{\mu\nu}$ is the usual single-nucleon tensor appearing in standard PWIA for
outgoing nucleon polarized scattering, while ${\cal Z}^{\mu\nu}$ and ${\cal N}^{\mu\nu}$ are
new single-nucleon tensors that enter only when the bound nucleon wave functions contain non-zero
negative-energy projections. Note also that the three single-nucleon tensors in Eqs.~(\ref{eq23}-\ref{eq25}) can be
split into symmetric and antisymmetric terms. The symmetric terms do not depend on the recoil
nucleon polarization and, when contracted with the symmetric part of the leptonic tensor which
does not depend on the electron polarization, give rise to the four unpolarized responses
$R^L_0$, $R^T_0$, $R^{TL}_0$ and $R^{TT}_0$ studied in detail in~\cite{Cab98a} (note that 
the unpolarized responses and single-nucleon tensors in~\cite{Cab98a} are twice the ones given 
here because of the spin projection operator included for the outgoing nucleon polarized 
situation; i.e., the factor of two is
recovered upon performing ``sum over final"). 
In contrast, the antisymmetric terms in Eqs.~(\ref{eq23}-\ref{eq25}) depend linearly on the outgoing
nucleon four-spin $S_N^\mu$. Their contractions with the antisymmetric part of the leptonic
tensor, which contains the dependence on the incident electron polarization, give rise to the 
positive- and negative-energy contributions in the spin-dependent responses $R^{T'}_{l,s}$ and 
$R^{TL'}_{l,s,n}$.

Next, let us recall that using the spin precession technique presented in~\cite{Cab93},
the non-diagonal spin single-nucleon tensor ${\cal N}^{\mu\nu}$, can be written in terms of 
a diagonal tensor constructed from spinors quantized with respect to a spin axis pointing
along a generic direction, ${\cal R}^{\mu\nu}(\theta_R,\phi_R)$, as
\be
{\cal N}^{\mu\nu}=\cos\theta {\cal R}^{\mu\nu}(0,0)+
                \sin\theta\left(\cos\phi {\cal R}^{\mu\nu}(\frac{\pi}{2},0)+
                                \sin\phi {\cal R}^{\mu\nu}(\frac{\pi}{2},\frac{\pi}{2})\right).
\label{eq26}
\ee
Here $\theta$, $\phi$ are the angles defining the direction of the bound nucleon momentum $\np$ and
the tensor ${\cal R}^{\mu\nu}$ is given as
\be
{\cal R}^{\mu\nu}(\theta_R,\phi_R)=\frac{1}{8M_N}Tr\left[
        \slslash \overline{J}^{\mu}(1+\gamma _{5}\snslash)(\pnslash+M)J^{\nu}\right] \, ,
\label{eq27}
\ee
which is linear in the bound nucleon spin four-vector $S_L^\mu$; its antisymmetric part also depends linearly on the
spin four-vector of the outgoing nucleon $S_N^\mu$. The angles $\theta_R$ and $\phi_R$ define the
direction of the bound nucleon spin $\ns_L$ in the frame in which the bound nucleon is at rest.
In Appendix B we show explicit expressions for the antisymmetric parts of the single-nucleon
tensors ${\cal W}^{\mu\nu}$, ${\cal Z}^{\mu\nu}$ and ${\cal R}^{\mu\nu}$
for different current operators. Explicit expressions for their symmetric parts were already
presented in~\cite{Cab98a}.

In analogy to the standard PWIA, where the cross section factorizes into a single-nucleon cross section
and the momentum distribution corresponding to a non-relativistic bound orbital, within RPWIA
one may also introduce single-nucleon cross sections $\sigma_{uu}^{eN}$, 
$\sigma_{vv}^{eN}$ and $\sigma_{uv}^{eN}$, which are constructed by contracting the single-nucleon
tensors ${\cal W}^{\mu\nu}$, ${\cal Z}^{\mu\nu}$ and ${\cal N}^{\mu\nu}$ appearing
in the three hadronic tensor contributions in Eqs.~(\ref{eq15}-\ref{eq17}) with the
leptonic tensor $\eta_{\mu\nu}$ (see~\cite{Cab98a} for details).
Then the differential cross section within RPWIA can be written as
\be
\frac{d\sigma}{d\varepsilon'd\Omega_{e}d\Omega_{N}}
=\frac{p_{N}M_N M_{B}}{M_{A}f_{rec}}\left[\sigma^{eN}_{uu}N_{uu}(p)
        +\sigma^{eN}_{vv}N_{vv}(p)+\sigma^{eN}_{uv}N_{uv}(p)\right] \, ,
\label{eq31}
\end{equation}
where $\sigma^{eN}_{uu}$ is the free polarized electron-nucleon cross section already appearing
in the standard PWIA analysis. Here $\sigma^{eN}_{vv}$ and $\sigma^{eN}_{uv}$ are new
components that arise only because of the negative-energy projections and that may only appear
in scattering from a bound nucleon. As already mentioned, this
constitutes an important difference between PWIA and RPWIA.
Whereas in PWIA the differential cross section factorizes into
two terms, the electron-nucleon cross section and the spectral function, 
in RPWIA the differential cross section depends on both positive- and 
negative-energy projections of the relativistic bound nucleon wave function. This result breaks the factorization property in the sense described in PWIA, namely, a clear separation into
two terms, one describing the electron-nucleon scattering
and the other connected to the nuclear structure of the target. 

Within RPWIA one may also use the decomposition of currents into longitudinal and transverse
components and then introduce the single-nucleon responses
\( {\cal R}_{uu}^{k,k'},{\cal R}_{vv}^{k,k'}\) and 
\({\cal R}_{uv}^{k,k'} \) that are given by taking the appropiate components of the single-nucleon
tensors \( {\cal W}^{\mu \nu }, {\cal Z}^{\mu \nu }\) and \({\cal N}^{\mu \nu } \), as is done
in PWIA. 
The hadronic response functions can be written in RPWIA in the form
\be
\label{eq:hadronresp}
R^{\alpha}=R^{\alpha}_{P}+R_{N}^{\alpha}+R^{\alpha}_{C}\, ,
\end{equation}
where the three components are given by
\begin{equation}
\label{eq:hadronpos}
R_{P}^{\alpha}={\cal R}^{\alpha}_{uu}N_{uu}(p)
\end{equation}
\begin{equation}
R_{N}^{\alpha}={\cal R}^{\alpha}_{vv}N_{vv}(p)
\label{eq37}
\end{equation}
\begin{equation}
R_{C}^{\alpha}={\cal R}^{\alpha}_{uv}N_{uv}(p)
\label{eq38}
\end{equation}
with \( \alpha = L,T,TL,TT,T',TL' \). In standard PWIA only the responses
${\cal R}^\alpha _{uu}$ occur. 


To finish the discussion in this section, let us present some brief comments on the
transferred polarization $\nP'$. This is basically given as the ratio between the recoil nucleon
polarized responses and the unpolarized responses. In particular, their three components
$P'_l$, $P'_s$ and $P'_n$ are given as
\ba
P'_{l,s} &=& \frac{\left(v_{T'}R^{T'}_{l,s}+v_{TL'}R^{TL'}_{l,s}\right)\cos\phi}
                {v_LR^L_0+v_TR^T_0+v_{TL}R^{TL}_0\cos\phi+v_{TT}R^{TT}_0\cos 2\phi} 
\label{eq39} \\
P'_{n} &=& \frac{v_{TL'}R^{TL'}_{n}\sin\phi}
                {v_LR^L_0+v_TR^T_0+v_{TL}R^{TL}_0\cos\phi+v_{TT}R^{TT}_0\cos 2\phi} \, .
\label{eq40}
\ea
Note that $P'_n$ only contributes for out-of-plane kinematics. The transferred polarization,
as a ratio observable, is very well-suited in experimental studies to
minimizing systematic errors. Different aspects of the underlying
dynamics can be revealed when studying such observables than when
considering the responses themselves; for instance, concepts such as
the overall normalization of the cross section and responses connects
to issues of spectroscopic factors, and such normalizations
largely drop out when forming ratios of observables. Thus, such
transferred polarizations hold promise for shedding light on other
aspects of the problem such as the roles played by FSI and by both
kinematical and dynamical relativistic effects.
A full RDWIA analysis is surely needed before general remarks
on the sensitivity of $\nP'$ to FSI could be presented and work along these lines is presently in
progress~\cite{Cris3}. 


\section{Analysis of the results}


In this section we present the results obtained as functions of the missing momentum $p$,
for the various recoil nucleon
polarization observables corresponding to proton knockout from the $p_{1/2}$ and $p_{3/2}$ 
shells in $^{16}$O leading to the residual nucleus $^{15}$N. Four different kinematical situations have been selected:
\begin{enumerate}
\item $q=500$ MeV/c, $\omega=131.56$ MeV;
\item $q=1$ GeV/c, $\omega=445$ MeV;
\item $\theta_N=0$, $p_N=490$ MeV/c;
\item $\theta_N=0$, $p_N=1$ GeV/c.
\end{enumerate}
Kinematics 1 and 2 correspond to $(q,\omega)$-constant kinematics (sometimes also referred as quasi-perpendicular
kinematics). In both cases the value selected of the transfer energy $\omega$ corresponds almost to 
the quasielastic peak value (the scaling variable $y$ is equal to zero in this situation), where one expects the validity of the impulse approximation in describing
the electron scattering process to be the highest. The interest in selecting
these two kinematics is twofold. First, the dynamical relativistic effects for 
the unpolarized responses have been already analysed in~\cite{Cab98a} for kinematics 1. 
Secondly, kinematics 2 roughly corresponds to that used in the 
experiments E89-003~\cite{E89003} and E89-033~\cite{E89033} performed at JLab. 
Kinematics~3 and 4 correspond to parallel kinematics where the recoil nucleon
momentum $\np_N$ is detected parallel to the transfer momentum $\nq$. The kinetic energy of the
outgoing nucleon is also fixed, as well as the electron beam energy.
Within parallel kinematics let us recall that positive
values of the missing momentum $p$ correspond to $\theta=0$, i.e.,
$\np$ and $\nq$ are parallel, whereas negative $p$-values correspond to 
$\theta =\pi$, i.e., $\np$ and $\nq$ antiparallel.

It is important to remark that
each of the types of kinematics, $(q,\omega)$-constant and parallel, presents clear differences that may lead
to significant effects in the polarization observables. Within $(q,\omega)$-constant kinematics, varying
the missing momentum $p$ means changing the direction in which the outgoing nucleon is detected. 
The kinetic energy of the outgoing nucleon should also vary to fulfill energy conservation, although
this variation is negligible in most of the cases (see discussion in~\cite{Edu95}). 
Within parallel kinematics, varying the missing momentum $p$ means changing the value of the
transfer momentum $q$. Although the transfer energy $\omega$ also depends on $p$, its variation
in the range, $-200\leq p\leq 300$ MeV/c,
is almost negligible since $M_B >> p$, meaning that as $p$ varies one is moving far away from the
quasielastic peak (QEP). This is in contrast with the previous kinematics 
where one is always at the center of the QEP. A very general and systematic study of the
different ingredients that enter in the analysis of the kinematics of coincidence electron
scattering reactions has been presented in~\cite{Cris1}.

As has been discussed at length in previous work~\cite{Cab98a,Cab93,For83}, the choice of the
nucleon current operator $J^\mu$ that enters in the description of $(e,e'N)$ processes
is to some extent arbitrary. In this work we 
have considered the two choices of the current operator known as $CC1$ and $CC2$.
Moreover, 
Moreover, the current is not conserved and different gauges should be considered:
i) Landau ($NCC1$, $NCC2$) - no current conservation imposed, ii) Coulomb ($CC1^{(0)}$,
$CC2^{(0)}$) - current conservation imposed by eliminating the third component, and
iii) Weyl ($CC1^{(3)}$, $CC2^{(3)}$) - current conservation imposed by eliminating the time
component. 
This subject is studied in detail in Section 3.1 where we show the behaviour of the polarized
single-nucleon responses insofar as their dependence on the 
current-conservation prescription selected is concerned. 
We present results for the positive-energy, as well as for 
the negative-energy components. 

In Sections 3.2 and 3.3 we present the results for the polarized hadronic responses and transferred
polarization asymmetries, respectively. The momentum distribution components in Eqs.~(\ref{eq18}-\ref{eq20})
were studied in detail in~\cite{Cab98a,Cab98b} for the $p_{1/2}$ and $p_{3/2}$ shells
in $^{16}$O. As shown there, the positive-energy projections are clearly dominant at low $p$, while
for $p>300$ MeV/c the negative-energy projections start to play an important role and the differences
between off-shell prescriptions are enhanced. Moreover, the behaviour of the dynamical enhancement
function $\beta_\kappa$ in Eq.~(\ref{eq22}) for stretched ($p_{3/2}$) and jack-knifed ($p_{1/2}$) states
is quite different. This explains why the amplitudes of the negative-energy projections for 
the $p_{1/2}$ state are much larger than those for the $p_{3/2}$.
All of the results presented in Sections 3.2 and 3.3 correspond to a bound state wave function computed
within the Walecka relativistic model~\cite{SeWa86} using the parameters of the set 
HS~\cite{Hor81,Hor91}. Results
with the NLSH~\cite{Sharma93} and NL3~\cite{Lalaz97} sets are quite
similar and a discussion of those would follow the same trends.


\subsection{Polarized single-nucleon responses}


In Figs.~\ref{sec31fig1} and~\ref{sec31fig3} we show the behaviour of the recoil 
nucleon polarized single-nucleon responses. 
For brevity in this work we consider only coplanar kinematics, i.e., $\phi =0$, 
which means that only four
recoil nucleon polarized responses are accesible, $R^{T'}_{l,s}$ and $R^{TL'}_{l,s}$.
For each single-nucleon response we
show the results for the positive-energy $uu$ (left-hand panels), and negative-energy $uv$ (middle panels) and $vv$
(right-hand panels) contributions. Curves for the six off-shell prescriptions
are shown in all of the cases. 

\subsubsection*{\bf Analysis in $(q,\omega)$-constant kinematics}

\begin{figure}
{\par\centering \resizebox*{0.6\textwidth}{0.45\textheight}{\rotatebox{270}{\includegraphics{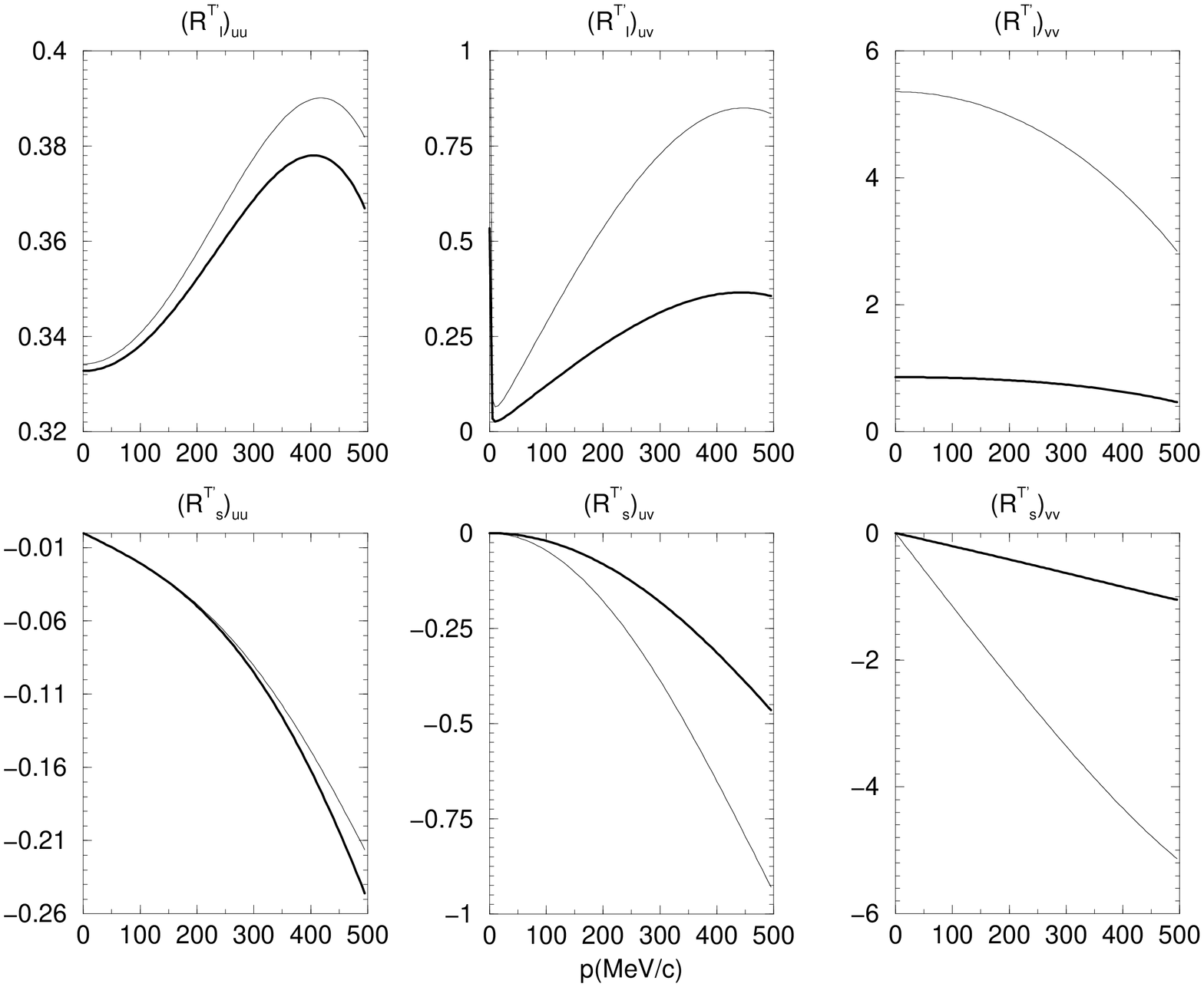}}} \par}
{\par\centering \resizebox*{0.6\textwidth}{0.45\textheight}{\rotatebox{270}{\includegraphics{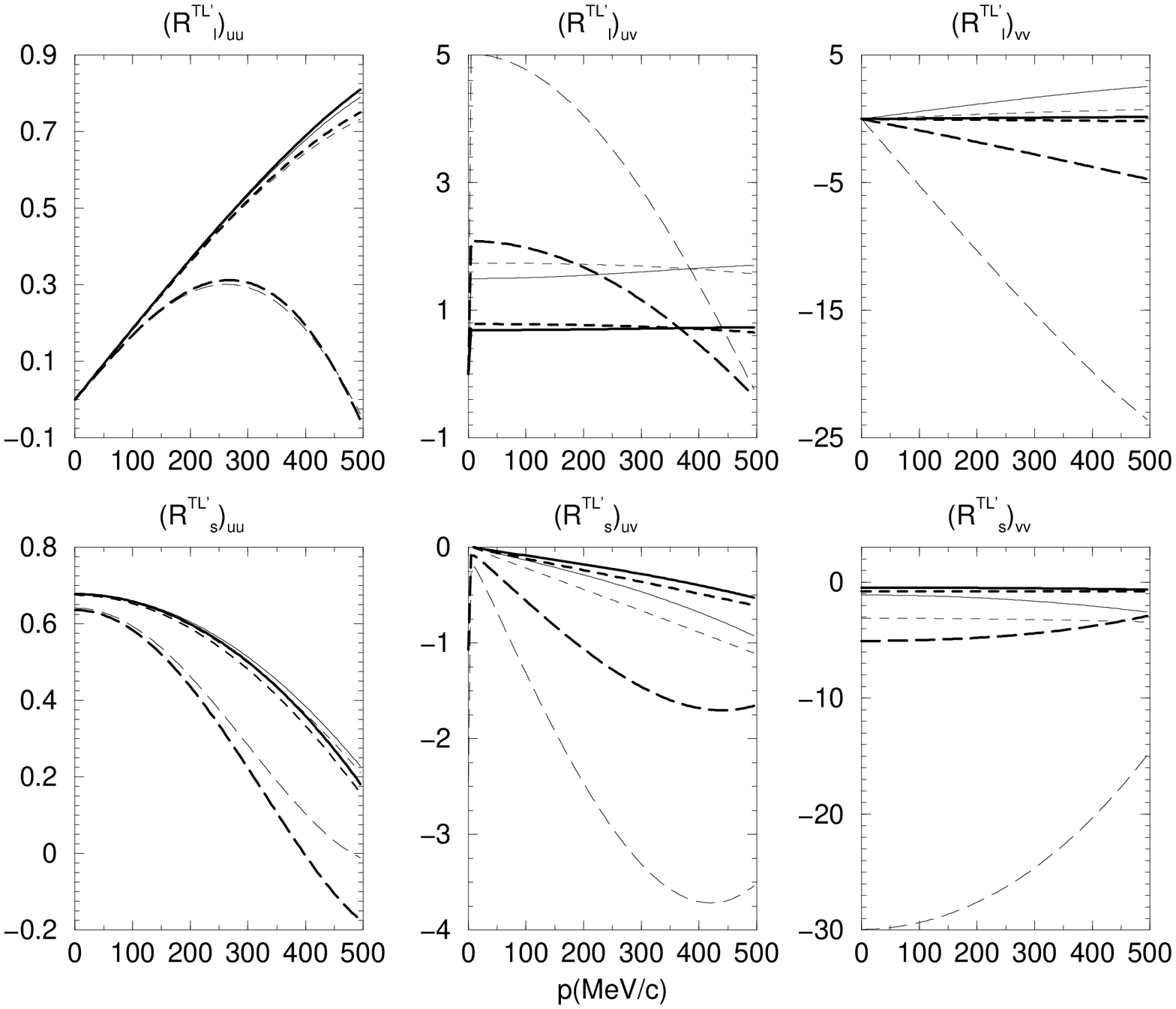}}} \par}
\caption{\label{sec31fig1}Polarized single-nucleon response functions for kinematics 1
(see text).
The two top panels correspond to the purely transverse $T'$ responses, while
the two bottom ones show results for $TL'$ responses. Thick lines correspond
to prescriptions based on the $CC2$ current operator and thin lines to the $CC1$
current. Results are shown for Landau ($NCC1/NCC2$) (solid lines),
Coulomb (\protect\(CC1 ^{(0)}\protect \)/\protect\(CC2 ^{(0)}\protect \)) (short-dashed lines) and
Weyl (\protect\(CC1 ^{(3)}\protect \)/\protect\(CC2 ^{(3)}\protect \)) (long-dashed lines) gauges.}
\end{figure}


We start our discussion with the case of $(q,\omega)$-constant kinematics. For simplicity we
only show results corresponding to kinematics 1 (Fig.~\ref{sec31fig1}).
The analysis of kinematics 2 follows in general similar trends. 
Let us consider first the
two purely transverse polarized responses ${\cal R}^{T'}_{l,s}$. These responses
only depend on the current operator ($CC1$ vs $CC2$) selected, i.e., they are not affected by 
gauge ambiguities. Fig.~\ref{sec31fig1} shows that, while the positive-energy components 
$({\cal R}^{T'}_{l,s})_{uu}$ obtained with the two currents are rather similar, the negative-energy
contributions $({\cal R}^{T'}_{l,s})_{uv}$ and $({\cal R}^{T'}_{l,s})_{vv}$ deviate significantly,
especially for the latter. This behaviour was already observed for the
purely transverse unpolarized single-nucleon responses ${\cal R}^T$ and ${\cal R}^{TT}$ 
(see discussion in~\cite{Cab98a}).
Although not shown, the same qualitative behaviour is found for kinematics 2.

Comparing the results obtained for the negative-energy components with the positive-energy ones, we
observe that the $uv$ and $vv$ contributions are clearly maximized by
the $CC1$ current operator. This is also consistent with the results obtained for the
unpolarized responses and reflects a stronger dependence in general on the negative-energy
projections $\beta_\kappa$ when using the $CC1$ operator. It is also interesting
to observe the different contributions of the $uv$ and $vv$ components in
both transverse responses ${\cal R}^{T'}_l$ and ${\cal R}^{T'}_s$. The negative-energy
($uv$ and $vv$) contributions compared with the positive-energy ($uu$) one are considerably larger for
${\cal R}^{T'}_s$. This result will be discussed further in next
section, when we show the total hadronic responses. 

The interference $TL'$ polarized responses are shown in the two bottom panels of 
Fig.~\ref{sec31fig1}. In this case, the responses depend on the gauge and hence
six different curves corresponding to the six `off-shell' prescriptions considered are presented. 
From these, the
$CC1^{(3)}$ and $CC2^{(3)}$ choices cause large deviations in all of the components, including the
positive-energy ones. On the contrary, the four remaining prescriptions give very similar
results for the $uu$ component, deviating clearly in the case of the negative-energy components,
particularly when comparing results corresponding to prescriptions using the two current
operators $CC1$ and $CC2$. In analogy with the two $T'$ responses, the $CC1$ choice maximizes 
the `off-shell' ambiguities, as well as the contribution
of the $uv$ and $vv$ components compared with the positive-energy one. Finally, note that the
relative contribution of the negative-energy components is more important for the
${\cal R}^{TL'}_l$ response. As will be shown in next section, this means that the hadronic
response $R^{TL'}_l$ can be more sensitive to dynamical relativistic effects than
$R^{TL'}_s$. 

In~\cite{Cris1} we introduced the concept of class number and its connection with
the analysis of the kinematical relativistic effects. Here we extend that to
include also the $uv$-cross contributions ($C$), i.e., dynamical relativistic effects. To make
clearer the discussion that follows, let us rewrite the response functions according to
the generic form at low-$\chi$:
\begin{equation}
{\cal R} = \chi^n \alpha_0 [ 1 + \alpha_2 \chi^2 + {\cal O}(\chi^4) ],
\label{lowchi}
\end{equation}
where $\chi\equiv (p/M_N)\sin\theta$ 
and $n=0$, 1, 2 labels the class type ``0'', ``1'', ``2'', etc (see~\cite{Cris1} for
details).

\begin{table}
{\centering \begin{tabular}{ccccc}
\hline
Response & $\alpha_2^P$ & $\alpha_2^C$
\\
\hline
\hline
\\
$L$ & $1.4 (n=0)$ & $0.1 (n=1)$ \\
$T$ & $1.2 (n=0)$ & $0.1 (n=1)$ \\
$TL$ & $0.3 (n=1)$ &  $1.2 (n=0)$ \\
$TT$ & $0.1 (n=2)$ & $0.7 (n=1)$ \\
$T'_l$ & $1.1 (n=0)$ & $-0.3 (n=1)$ \\
$TL'_l$ & $-0.1 (n=1)$ & $0.2 (n=0)$ \\
$T'_s$ & $3.3 (n=1)$ & $0.004 (n=2)$ \\
$TL'_s$ & $-1.4 (n=0)$ & $0.2 (n=1)$ \\   
\hline 
\end{tabular}\par}
\caption{\label{classcoef} Coefficients $\alpha_2$ given by the decomposition in Eq.~(\ref{lowchi}) for the positive and crossed response functions.}
\end{table}

Truncating all expansions of the above type at terms involving $\alpha_2$ the results obtained
are presented in table~\ref{classcoef}. We compare the cross ($C$) contributions with the PWIA
results (labelled ``$P$''). In~\cite{Cris1} we have shown that
kinematical relativistic effects are typically
larger for class ``0'' responses than for reponses of classes ``1''
and ``2''. Now an extended pattern emerges for the terms of type ``$C$''
involving dynamical relativistic effects.
First, two groups occur, one in which the class
of the $C$ contribution is one higher than the class of the $P$
contribution ($L$, $T$, $T'_l$, $T'_s$ and $TL'_s$), and another in
which the class of the $C$ contribution is one lower than the
class of the $P$ contribution ($TL$, $TT$ and $TL'_l$). In the latter cases we expect to 
see greater consequences of dynamical relativity than in the former. 
Second, we see from above that the importance of higher-order
terms in the $\chi^2$ expansions for type $C$
typically correlates in the same way it does for the $P$ expansions:
the class ``0'' cases generally show larger effects than do the
responses in classes ``1'' and ``2''.

\subsubsection*{\bf Analysis in parallel kinematics}

\begin{figure}[t]
{\par\centering \resizebox*{0.6\textwidth}{0.45\textheight}{\rotatebox{270}
{\includegraphics{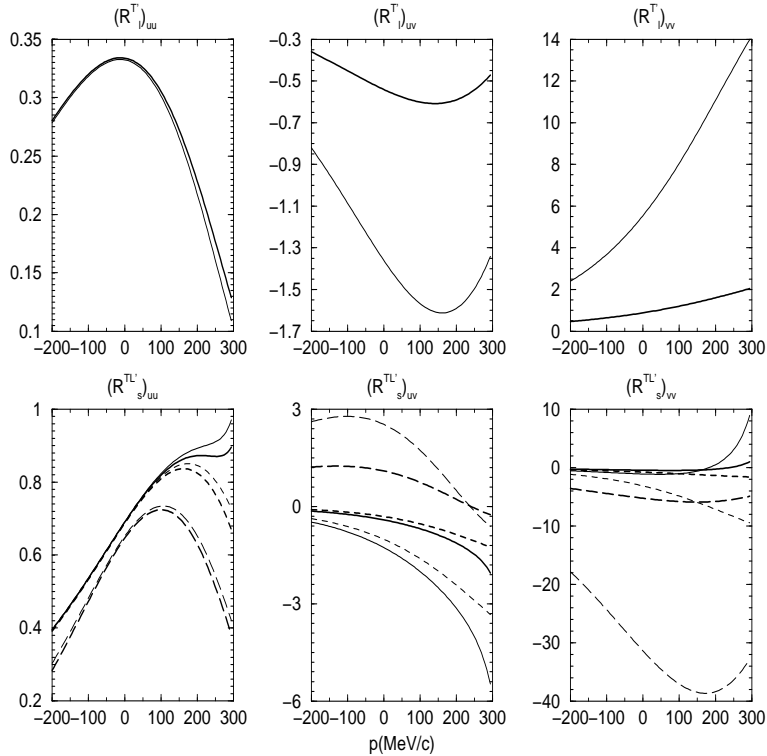}}} \par}
\caption{\label{sec31fig3} Polarized single-nucleon responses in parallel kinematics 3.
The labeling of the curves is as in Fig.~\ref{sec31fig1}.}
\end{figure}

Fig.~\ref{sec31fig3} shows the results for parallel kinematics where
only two polarized responses survive, 
namely ${\cal R}^{T'}_l$ and ${\cal R}^{TL'}_s$. In this case the positive- and negative-energy
components of the polarized response ${\cal R}^{T'}_l$ are proven to be identical to the
respective components of the unpolarized transverse response ${\cal R}_0^T$.
In fact, these two responses are identical in parallel kinematics 
for $j=1/2$ orbits even including FSI in the analysis
(see~\cite{RaDo89,Giusti89} for details). Here we only present
the results corresponding to kinematics 3. The discussion and analysis of responses for
kinematics 4 follow in general similar trends.
Some of the points discussed for the previous kinematics can
be also applied here. In particular, note that the two current operators give very similar results
for the positive-energy component of ${\cal R}^{T'}_l$, whereas they differ strongly for the
negative-energy contributions, being maximized for the $CC1$ current operator. In the case
of ${\cal R}^{TL'}_s$, the $CC1^{(3)}$ and $CC2^{(3)}$ prescriptions produce the biggest differences 
even for the $uu$ component. Again, choices based on the $CC1$ current maximize the role of the
negative-energy contributions. 

It is interesting to remark that the two single-nucleon responses surviving in parallel kinematics are the ones
that also show less sensitivity to dynamical relativistic effects in $(q,\omega)$-constant kinematics 
(Fig.~\ref{sec31fig1}).
However, by looking more carefully at the relative contributions to these two responses
of the negative-energy $uv$ and $vv$ components, the sensitivity is shown to be more significant in the
case of parallel kinematics. We will return to this issue in the next section when showing the behaviour
of the total hadronic responses.

To finish the discussion on parallel kinematics we should also point out the basic differences observed
in the results for positive and negative $p$-values. As explained above, positive (negative) $p$-values
mean parallel (antiparallel) direction between $\np$ and $\nq$. From the responses 
shown in Fig.~\ref{sec31fig3},
one observes that `off-shell' ambiguities are considerably reduced for $p$-negative. This fact
is maintained for the positive-energy, as well as for the negative-energy components. One can also
observe that the
relative contribution of $uv$ and $vv$ components, compared with $uu$, although important 
is also reduced in the case of negative $p$-values. From these results, it is clear
that the negative-$p$ region
should be favoured in the experiments. This coincides with the conclusions
presented in~\cite{Udi00}. In general, response functions seem to be less
affected by `off-shell' uncertainties in that case (see~\cite{Cris1}), although relativistic dynamical effects are still important.


\subsection{Hadronic responses}


In this section we present results for the recoil nucleon
polarized hadronic responses as defined in Eq.~(\ref{eq:hadronresp}).
Our aim is to analyze the relative importance of the negative-energy projection contributions
for the various `off-shell' prescriptions considered. As already mentioned, we consider
the cases of the proton being knocked out from the $p_{1/2}$ and $p_{3/2}$ shells in $^{16}$O.

\begin{figure}
{\par\centering \resizebox*{0.5\textwidth}{0.45\textheight}{\rotatebox{270}
{\includegraphics{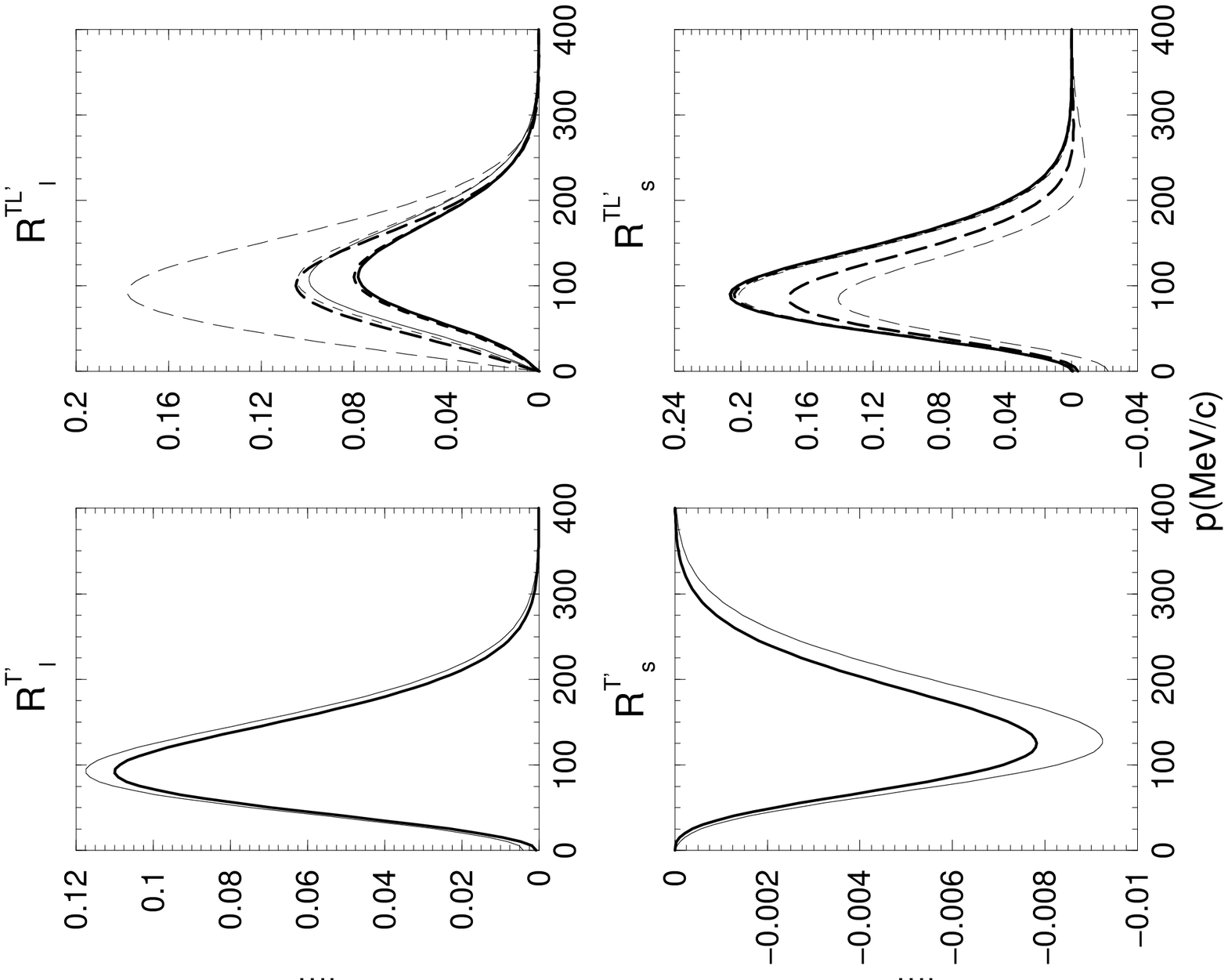}}} \par}
\vspace{0.5cm}
{\par\centering \resizebox*{0.5\textwidth}{0.45\textheight}{\rotatebox{270}
{\includegraphics{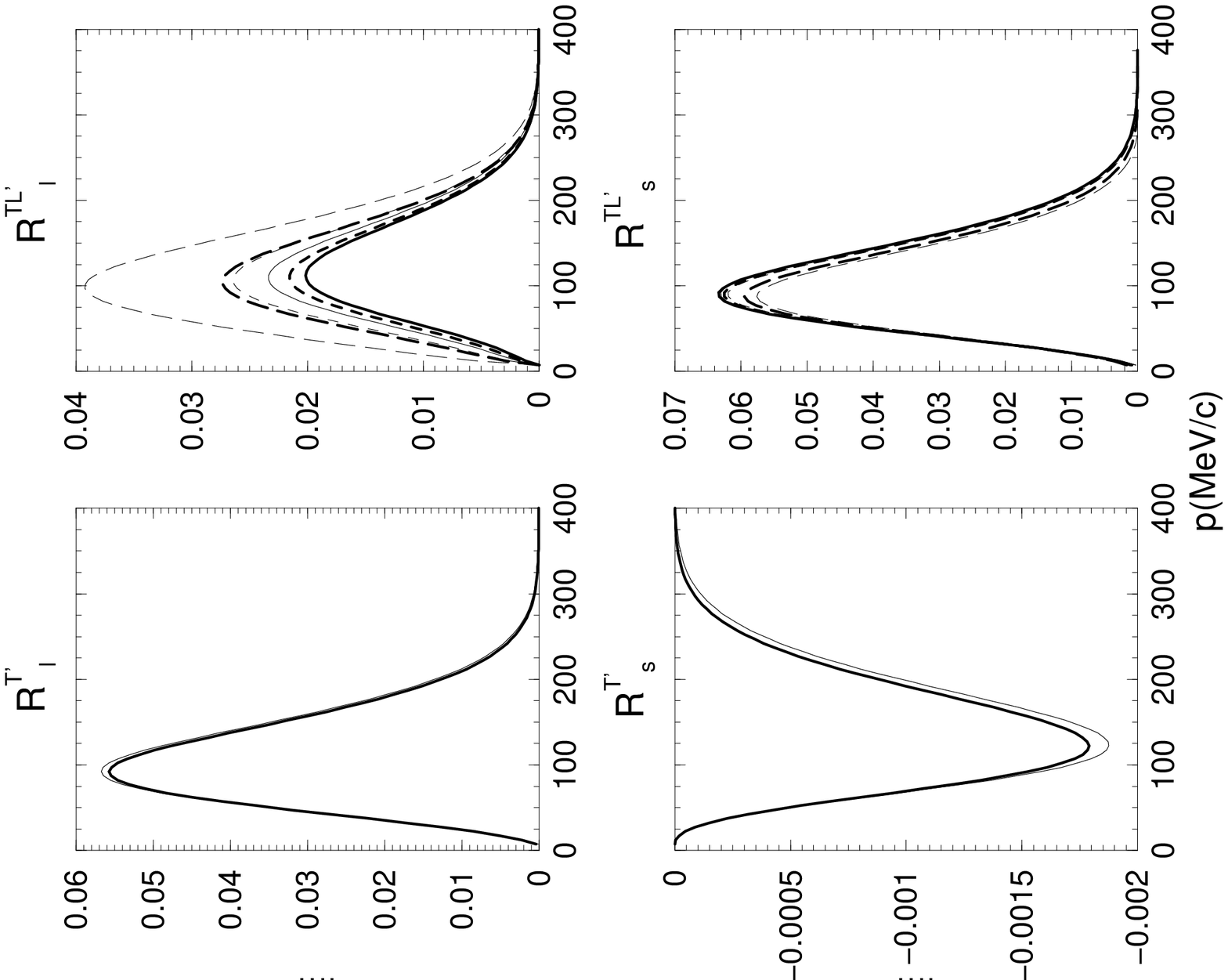}}} \par}
\caption{\label{sec32fig1}Polarized hadronic response functions for the 
\protect\( p_{1/2}\protect \)
shell in $(q,\omega)$-constant kinematics. Top panels correspond to kinematics 1 and bottom panels to
kinematics 2 (see text). The curves are labelled as in Fig.~\ref{sec31fig1}.}
\end{figure}

\subsubsection*{\bf Analysis in $(q,\omega)$-constant kinematics}

In Fig.~\ref{sec32fig1} we show the total hadronic responses for $p_{1/2}$ and 
$(q,\omega)$-constant kinematics. The top panels correspond to kinematics 1 and the bottom panels to
kinematics 2. As we see, apart from quantitative details, the behaviour of the responses is
quite similar for the two choices of kinematics, and accordingly the discussion that follows applies to both.
Concerning the purely transverse $T'$ responses, we observe that the ambiguity introduced by the
choice of current operator at the maxima of the responses ($p\sim100$ MeV/c, as our results are obtained for p-waves), although visible in the two responses, is larger in the
case where the spin polarization lies in the sideways direction (for the $R^{T'}_s$ response it is $\sim$16\% for kinematics 1 and $\sim$4\% for kinematics 2, while for $R^{T'}_l$ is $\sim$7\% for kinematics 1 and $\sim$2\% for kinematics 2). Note however, that in absolute
value $R^{T'}_s$ is more than one order of magnitude smaller than $R^{T'}_l$. The reason for the 
`off-shell' uncertainty in these responses can be traced back to the influence of the negative-energy projections of the bound nucleon wave function, given by the contribution of the crossed and negative components introduced in Eq.~(\ref{eq:hadronresp}). In fact, although not shown in the graphs, we have checked that for the positive terms there is almost no difference between
the results provided by the two current operators selected, whereas the crossed and negative terms strongly depend on the current choice, explaining the differences observed in 
Fig.~\ref{sec32fig1}. The relative influence introduced by these terms
at the maxima is $\sim$10\% ($\sim$4\%) 
for $CC1$ ($CC2$) choices in $R^{T'}_l$, and
$\sim$30\% ($\sim$11\%) for $CC1$ ($CC2$) in $R^{T'}_s$ in the case of kinematics 1, and
$\sim$3\% ($\sim$1\%) for $CC1$ ($CC2$) in $R^{T'}_l$, and
$\sim$9\% ($\sim$4\%) for $CC1$ ($CC2$) in $R^{T'}_s$ in the case of kinematics 2. These numbers explain not only why the `off-shell' ambiguity at the maxima of the responses is larger for $R^{T'}_s$ than for $R^{T'}_l$, but also why it is significantly reduced in the case of kinematics 2. We have also considered higher
values of $q$ to verify that as $q$ increases off-shell effects diminish.

The polarized interference transverse-longitudinal responses $R^{TL'}_l$ and $R^{TL'}_s$ are shown
for the six `off-shell' prescriptions in the right-hand panels of Fig.~\ref{sec32fig1}. It is
interesting to note the different behaviour shown by the various
prescriptions in the two responses.
For $R^{TL'}_s$ only the Weyl gauge prescriptions and particularly the
$CC1^{(3)}$ case produce significant
differences (larger for kinematics 1) with respect to the other prescriptions.
Results corresponding to Landau and Coulomb gauges with $CC1$ and $CC2$ currents are
very similar. As before, these facts are directly associated with the
role that the negative-energy 
projection components play in each case. For Landau and Coulomb gauge
prescriptions the role 
played by the NEP in this response is less than $\sim$4\%. In contrast, the
negative-energy contribution for the Weyl prescriptions are $\sim$28\% for $CC1^{(3)}$ and
$\sim$10\% for $CC2^{(3)}$ for kinematics 1, and $\sim$8\% for $CC1^{(3)}$ and
$\sim$3\% for $CC2^{(3)}$ for kinematics 2. However, it must be
pointed out that this choice of gauge leads to significantly different
results even in the positive components.

The case of the $R^{TL'}_l$ response is clearly different. Here the
spread amongst the six
`off-shell' prescriptions is wider. Not only does the Weyl gauge give rise to different results, but also
the responses corresponding to the two Landau gauge prescriptions ($NCC1$, $NCC2$) differ from the
ones corresponding to Coulomb ($CC1^{(0)}$, $CC2^{(0)}$). Interestingly, once one has selected the
current operator, the results obtained with both Coulomb and Landau gauges are very similar
in the case of kinematics 1, while the differences between them increase for kinematics 2, particularly for $CC1$ current.
Moreover, the effects introduced by the current choice for the Coulomb and Landau gauges, although 
visible, are much smaller than in the case of 
the Weyl gauge where the difference between
$CC1^{(3)}$ and $CC2^{(3)}$ amounts to a factor $\sim 1.7$ (kinematics~1) 
and $\sim 1.4$ (kinematics~2).
All of these effects in $R^{TL'}_l$ can be traced back to the behaviour of its various positive- and negative-energy components. The contributions given by the six prescriptions in the
crossed components are similar or even considerably larger ($CC1^{(3)}$) than the ones corresponding to the 
positive-energy terms --- from $\sim$35\% ($NCC2$, $CC2^{(0)}$) up to $\sim 250\%$ ($CC1^{(3)}$)
for kinematics~1, and from $\sim$35\% ($NCC2$) up to $\sim$180\% ($CC1^{(3)}$) 
for kinematics~2.
Moreover, the spread of the `off-shell' ambiguities
in the negative-energy terms is much wider than in the positive-energy contribution. 
These facts explain on one hand why the total $R^{TL'}_l$ response is much bigger than its purely
positive-energy contribution and on the other why the off-shell uncertainties are so large.

Summarizing, we conclude that the polarized response $R^{TL'}_l$ seems to present the highest
sensitivity to the relativistic dynamical enhancement of the lower components of the bound
nucleon wave function. Unfortunately, the `off-shell' uncertainty is consequently the largest.
The transverse response $R^{T'}_s$ also shows significant sensitivity to dynamical relativistic effects; however it is very small and hence difficult to measure. Finally, $R^{TL'}_s$ and $R^{T'}_l$ show less sensitivity
to the negative-energy contributions, and present the lowest spread
due to `off-shell' uncertainties. At the maxima of the responses, the off-shell ambiguity
in the two purely transverse responses $R^{T'}_{l,s}$
is significantly reduced for increasing $(q,\omega)$-values (within the QEP).
Concerning the interference $R^{TL'}_{l,s}$ responses, the off-shell effects
corresponding to the Landau and Coulomb gauges
remain quite similar when varying $q$ (under quasielastic conditions).

\begin{figure}
{\par\centering \resizebox*{0.5\textwidth}{0.45\textheight}{\rotatebox{270}
{\includegraphics{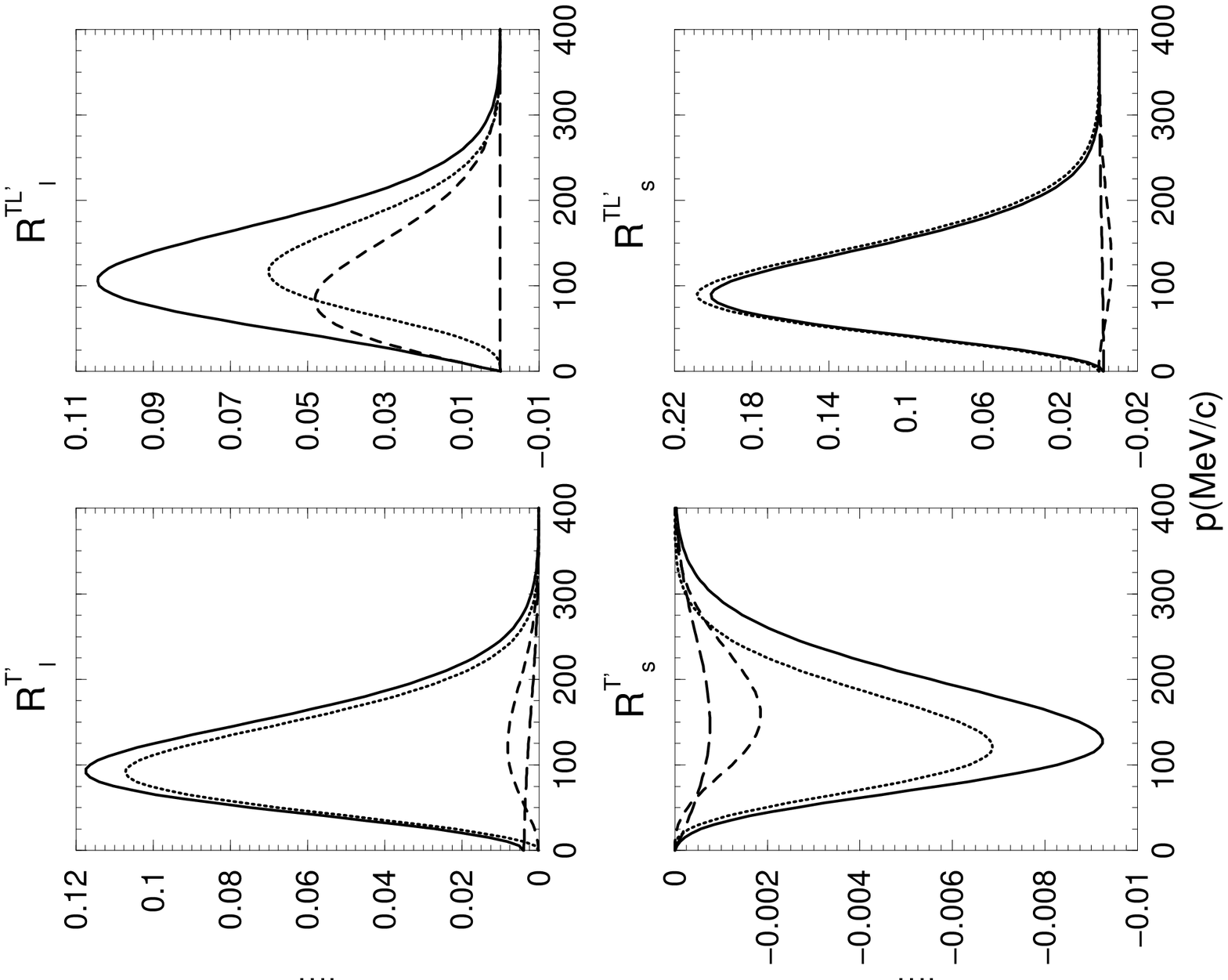}}} \par}
\vspace{0.5cm}
{\par\centering \resizebox*{0.5\textwidth}{0.45\textheight}{\rotatebox{270}
{\includegraphics{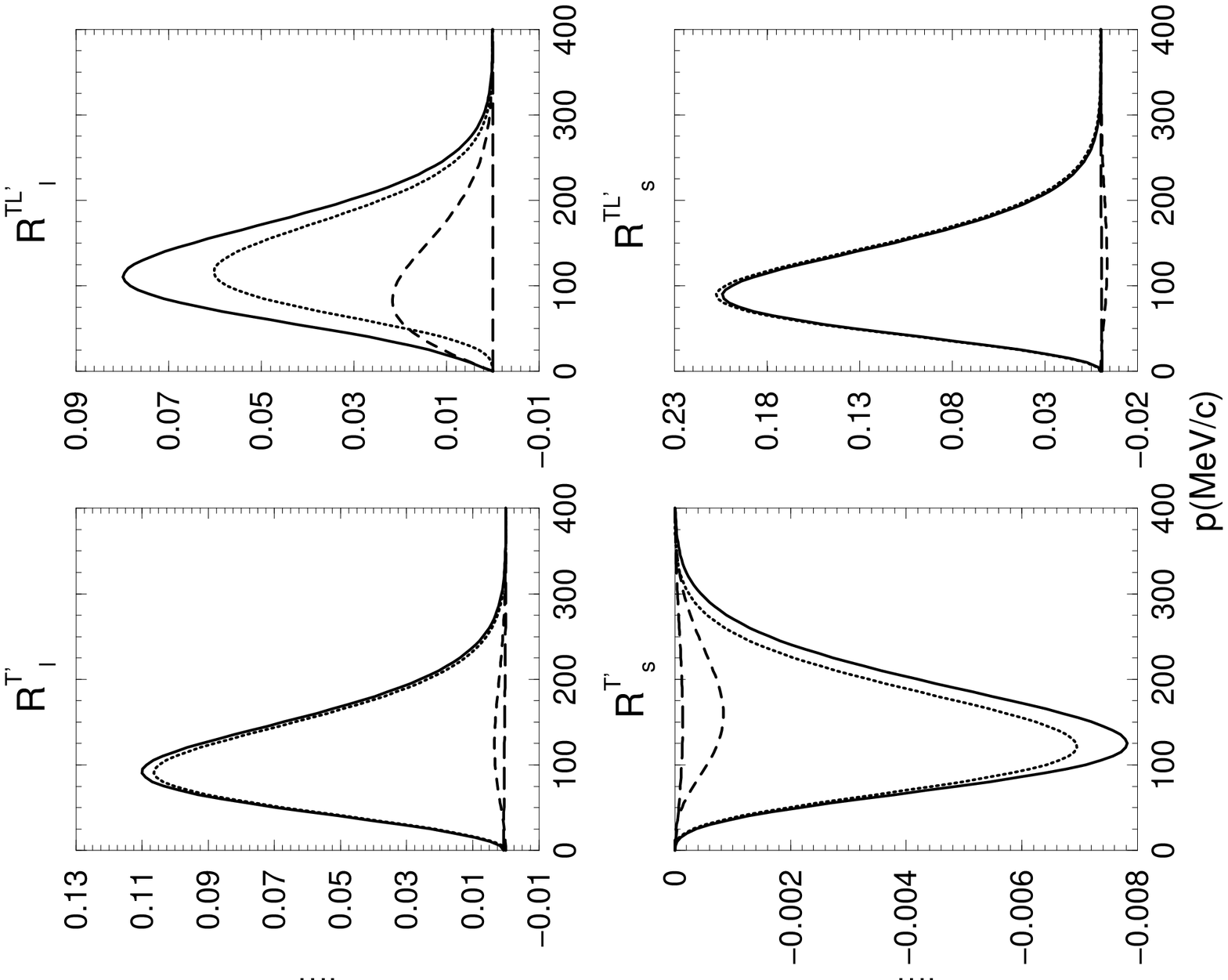}}} \par}
\caption{\label{sec32fig4}Hadronic response functions for the \protect\( 1p_{1/2}\protect \)
shell and kinematics 1. Top panels correspond to the \protect\( CC1^{(0)}\protect \) prescription
and bottom panels to \protect\( CC2^{(0)}\protect \). The fully-relativistic
response (solid line) is compared with the three components: $R^\alpha_P$
(dotted line), $R^\alpha_C$ (short-dashed line) and $R^\alpha_N$ (long-dashed line).}
\end{figure}


The role played by the negative-energy projection components in the four polarized
responses is more clearly seen in Fig.~\ref{sec32fig4}, where we present
the results corresponding to the Coulomb
gauge using the $CC1$ and $CC2$ choices of the current and
kinematics 1.
Results for the Landau gauge are seen to be very similar. 
We show the fully-relativistic result (solid) versus its three 
contributions as given by Eqs.~(\ref{eq:hadronpos}-\ref{eq38}): $R^\alpha_P$ (dotted),
$R^\alpha_C$ (short-dashed) and $R^\alpha_N$ (long-dashed). Note that
$R^\alpha_C$ and $R^\alpha_N$ both come from the negative-energy projection contributions.
The dynamical relativistic effects are easily appreciated by just comparing the
$N$ and $C$ terms with the total response.
In the case of the $CC1^{(0)}$ prescription 
(two top panels), it is clearly observed that in
two responses, $R^{T'}_l$ and $R^{TL'}_s$, the contribution 
of the negative-energy projections is almost negligible, that is, dynamical relativistic effects 
from the bound nucleon wave function do not significantly affect these responses. 
On the contrary, the two remaining polarized responses, $R^{T'}_s$ and $R^{TL'}_l$, are much more
sensitive. In both cases, although
the $N$ term does not contribute significantly to the total
result, the crossed term $C$ plays an important role, particularly
for the $R^{TL'}_l$ response where, as already discussed, its contribution is similar to the
one coming from the positive-energy projection.
This result resembles what appeared for the unpolarized interference
$TL$ response~\cite{Cab98a}. Hence there exists a strong discrepancy 
between RPWIA results and those corresponding to
the standard PWIA (we must recall that although the positive-energy term 
in Eq.~(\ref{eq:hadronpos}) is not identical to the PWIA result, for which 
we must take the non-relativistic momentum distribution $N_{nr}(p)$, 
the difference is very small provided that $N_{nr}(p)\sim N_{uu}(p)$).
As shown in the two bottom panels of Fig.~\ref{sec32fig4},
the results obtained for the Coulomb gauge with
the $CC2$ nucleon current operator follow a similar
trend to the ones discussed for the $CC1$ case, 
except for the magnitude of the relativistic effects. Although
the role of the $N$ and crossed $C$ terms is significantly reduced for
the $CC2$ current, their effects are still quite sizeable in $R^{T'}_s$ and $R^{TL'}_l$. 
This general behaviour is similar to the one already stated for the 
unpolarized responses in~\cite{Cab98a}. Finally, 
although not shown here for simplicity, we have also explored the results 
obtained for kinematics 2. 
We prove that dynamical relativistic effects in ${\cal R}^{TL'}_l$ are shown to be
independent of $q$, under QE conditions, when the $CC1^{(0)}$ prescription is used.
In the case of $CC2^{(0)}$ the role of the negative-energy components increases
slightly as $q$ goes to higher values.
On the contrary, for the two purely transverse responses $R^{T'}_{l,s}$, 
dynamical relativistic
effects are significantly reduced as $(q,\omega)$-values increase. 



In what follows we analyze the case of the $p_{3/2}$ shell and $(q,\omega)$-constant kinematics.
For simplicity
we only consider kinematics 1, the behaviour of the responses for kinematics 2 
being similar. We have checked (not shown here) that
the four hadronic responses obtained for the various off-shell prescriptions,
compared with the case of the $p_{1/2}$ orbit (Fig.~\ref{sec32fig1}), present a much smaller
`off-shell' uncertainty. This is connected with the minor
role played by the negative-energy projection components.
Despite the fact that the `off-shell' uncertainties for $R^\alpha_C$ and $R^\alpha_N$ in 
Eq.~(\ref{eq:hadronresp})
are of the same order of magnitude as in the case of the $p_{1/2}$ shell, their relative 
contribution in the total hadronic responses is much reduced;
only for the $R^{TL'}_l$ response is the contribution of the negative-energy projections significant.
This arises from the interestingly
different behaviour of the lower components in the two cases, namely,
that the $p_{1/2}$ case (``jack-knifed'') has lower components that
have an $s_{1/2}$ nature, whereas the $p_{3/2}$ case (``stretched'')
has lower components that go as $d_{3/2}$.

\begin{figure}[t]
{\par\centering \resizebox*{0.5\textwidth}{0.5\textheight}{\rotatebox{270}{\includegraphics{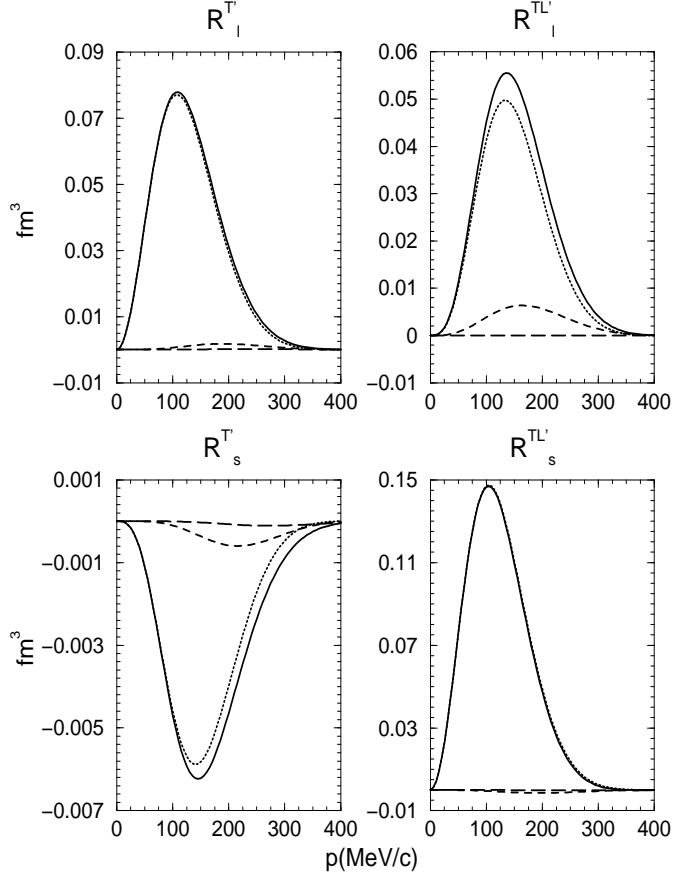}}} \par}
\caption{\label{sec32fig8}Hadronic response functions for the \protect\( 1p_{3/2}\protect \)
shell and kinematics 1 using the \protect\(CC1^{(0)}\protect \) prescription. The fully-relativistic
response (solid line) is compared with the three components: $R^\alpha_P$
(dotted line), $R^\alpha_C$ (short-dashed line) and $R^\alpha_N$ (long-dashed line).}
\end{figure}

The role of the negative-energy components is more 
clearly seen in Fig.~\ref{sec32fig8} where the terms $R^\alpha_P$, $R^\alpha_C$ and
$R^\alpha_N$ are plotted separately and compared with the
fully-relativistic results. The $CC1$ current, 
which magnifies dynamical relativistic effects, and the Coulomb gauge have been chosen. Note
that the positive contributions are very similar to the full responses. This result agrees with the
general study presented in~\cite{Cab98b} where the role of the lower components of the bound
nucleon wave function was investigated for different spin-orbit partner shells. In~\cite{Cab98b}
it was proven that the large deviation produced by different `off-shell' prescriptions in the
unpolarized responses $R^{TL}$ and $R^{TT}$ takes place for the jack-knifed states, but not for the
stretched states. This result, which also persists for the recoil nucleon polarized responses,
can be understood by the different behaviour of the dynamical
enhancement function $\beta_\kappa$ in Eq.~(\ref{eq22}) for the stretched and jack-knifed states.

\subsubsection*{\bf Analysis in parallel kinematics}

\begin{figure}[t]
{\par\centering \resizebox*{0.5\textwidth}{0.5\textheight}{\rotatebox{270}
{\includegraphics{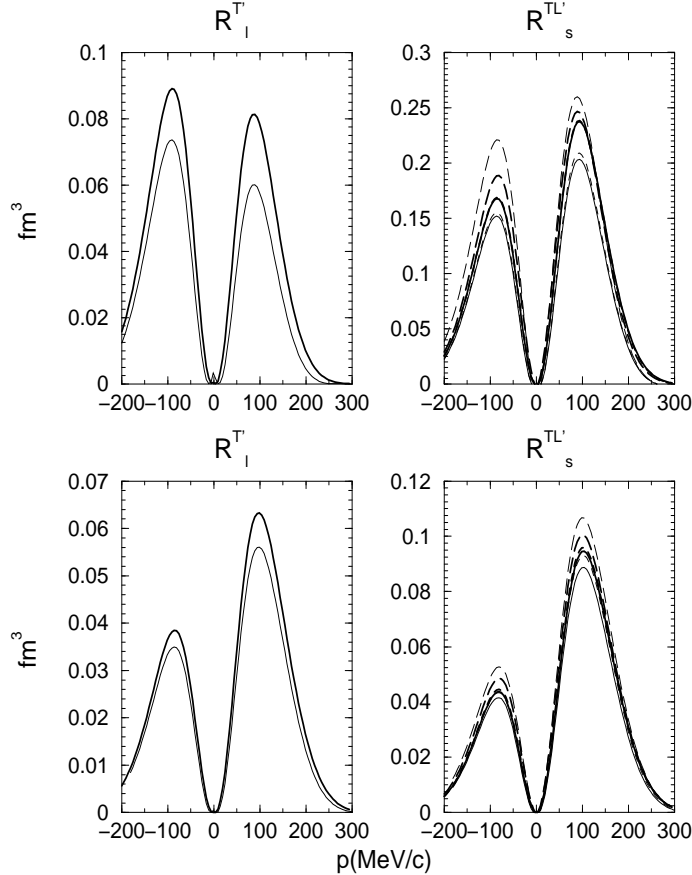}}} \par}
\caption{\label{sec32fig9}Polarized hadronic response functions for the \protect\( 1p_{1/2}\protect \)
shell in parallel kinematics. Top panels correspond to kinematics 3 and bottom panels to kinematics
4 (see text). The curves are labelled as in Fig.~\ref{sec31fig1}.}
\end{figure}

To finish the discussion in this section, we consider the case of parallel kinematics. 
We only show results for proton knock-out from the $p_{1/2}$ shell, as the discussion of
the $p_{3/2}$ orbit case follows similar trends to the ones already 
presented for $(q,\omega)$-constant kinematics.
The two surviving responses, $R^{T'}_l$ and $R^{TL'}_s$, are
shown in Fig.~\ref{sec32fig9} for the various `off-shell' prescriptions. The top panels correspond
to kinematics 3 and the bottom panels to kinematics 4. It is interesting to note that in the case of kinematics 4, i.e., higher outgoing nucleon kinetic
energy, the asymmetry in the responses for positive and negative $p$-values is 
significantly enhanced, although this result is already present at the level of the 
positive-energy contributions alone.
Concerning the purely transverse $T'$ response,
we have seen that the significant `off-shell' discrepancy observed at the maxima comes essentially from the crossed term, provided that both current operators $CC1$ and $CC2$ give the same positive-energy response. 
In contrast, the negative-energy contributions differ significantly, the $CC1$ result being much bigger in absolute value. Moreover, 
the ambiguity (evaluated at the maxima of the responses, $p\sim100$ MeV/c) 
introduced by the current choice is larger in the region of positive $p$-values, $\sim$27\% ($\sim$11\%) for kinematics~3 (kinematics~4), than for negative $p$, $\sim$17\% ($\sim$9\%) for kinematics~3 (kinematics~4). Note that off-shell uncertainties are also reduced when
the outgoing nucleon momentum is larger (kinematics~4). A similar
behaviour is also observed for the $R^{TL'}_s$ response in the case of the prescriptions based on
Coulomb and Landau gauges. The positive-energy contribution is
almost identical for these four `off-shell' prescriptions, whereas
significant differences appear in the negative-energy components, with the 
contributions based on the $CC1$ current choice being much more important. 
Moreover, this off-shell spread is larger within the
positive-$p$ region than for $p$ negative. 
In the case of the Weyl gauge, $CC1^{(3)}$ and $CC2^{(3)}$, the responses obtained
differ from the ones corresponding to the Coulomb and Landau gauges, even at the level of the 
positive-energy projection component. Also the behaviour of the
crossed term for the Weyl prescriptions is opposite (positive contribution) to the 
results of the other `off-shell' prescriptions. Moreover, whereas the Coulomb and Landau 
results in the crossed component are larger (in absolute value) for $p$-positive, the results
for the Weyl gauge are the opposite, i.e., larger contributions for $p$-negative.
These results are directly connected with the behaviour of the single-nucleon responses shown in
Fig.~\ref{sec31fig3}.

\begin{figure}[t]
{\par\centering \resizebox*{0.5\textwidth}{0.5\textheight}{\rotatebox{270}
{\includegraphics{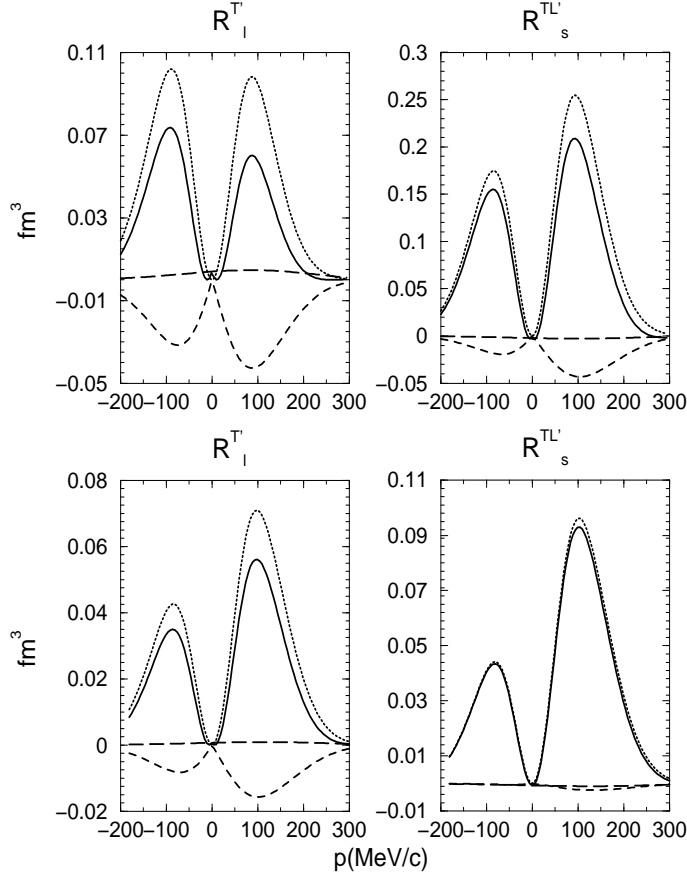}}} \par}
\caption{\label{sec32fig11} Hadronic response functions for the \protect\( 1p_{1/2}\protect \)
shell using the \protect\( CC1^{(0)}\protect \) prescription. The top panels correspond to kinematics 3 and the bottom ones to kinematics 4. The fully-relativistic
response (solid line) is compared with the three components: $R^\alpha_P$
(dotted line), $R^\alpha_C$ (short-dashed line) and $R^\alpha_N$ (long-dashed line).}
\end{figure}

As in the previous choices of kinematics, in order to clarify the role of the negative-energy projection components,
we show in Fig.~\ref{sec32fig11} the separate contribution of the terms $R^\alpha_P$, 
$R^\alpha_C$ and $R^\alpha_N$ for the Coulomb gauge and $CC1$ current choice. Results for Landau gauge
and the same current are similar, while choosing the $CC2$ current minimizes the negative-energy contributions. Comparing with the cases of ($q,\omega$)-constant 
kinematics (Fig.~\ref{sec32fig4}), we observe
that the two surviving responses in parallel kinematics present (at their peaks)
a stronger sensitivity to dynamical relativistic effects.
From these results it seems to be clear that choosing parallel kinematics
enhances the sensitivity to
dynamical relativistic effects in the two surviving polarized responses
compared with these same responses evaluated in $(q,\omega)$-fixed kinematics.
The reason for this can be traced back to the general discussion on kinematics presented at 
the beginning of Section~3.
Whereas for $(q,\omega)$-constant kinematics the $\omega$-value is chosen to fulfill the QEP condition,
in parallel kinematics, varying $p$ forces the transfer momentum $q$ also to vary, $\omega$ being practically constant. This means that one is moving away from the center 
of the QEP. As shown in~\cite{Cab98a} for the unpolarized responses, 
`off-shell' uncertainties and dynamical relativistic effects increase as one moves 
far away from the QEP. This fact could also explain why the sensitivity of the responses to
the negative-energy projections, evaluated at the maxima, is higher for kinematics 3:
$\sim$65\% vs $\sim$27\% in $R^{T'}_l$ and
$\sim$23\% vs $\sim$3\% in $R^{TL'}_s$ for $p=100$ MeV/c;
$\sim$37\% vs $\sim$21\% in $R^{T'}_l$ and
$\sim$12\% vs $\sim$1\% in $R^{TL'}_s$ for $p=-100$ MeV/c.
These results show that the dynamical enhancement of the 
lower components of the bound wave function have a greater effect in
the case of positive $p$-values. Note that the maxima
in the responses, located approximately at $p=\pm 100$ MeV/c, correspond to values of the transfer
momentum: $q\approx 400$ MeV/c ($q\approx 600$ MeV/c) for $p$-positive (-negative) and kinematics 3,
and $q\approx 900$ MeV/c ($q\approx 1100$ MeV/c) for $p$-positive
(-negative) and kinematics 4. In fact, we state without elaborating
further that the relative sizes of the effects seen
here appear to correlate with the value of $|y|/q$.


\subsection{Transferred Polarization Asymmetries}


In this section our aim is to analyze the behaviour of the transferred polarization 
asymmetries $P'_l$ and $P'_s$ introduced in Eq.~(\ref{eq39}).
We discuss the uncertainty introduced by the various off-shell prescriptions and
the role played by the dynamical enhancement of the lower components of the 
bound nucleon wave function. Let us recall that these transferred
polarization observables are thought to be very special in that they
have the potential to
provide information on the nucleon form factors in the nuclear medium; hence a careful analysis of the `off-shell'
and relativistic effects is crucial before one can hope to gain such insight.

\begin{figure}
{\par\centering \resizebox*{0.8\textwidth}{0.4\textheight}{\rotatebox{270}
{\includegraphics{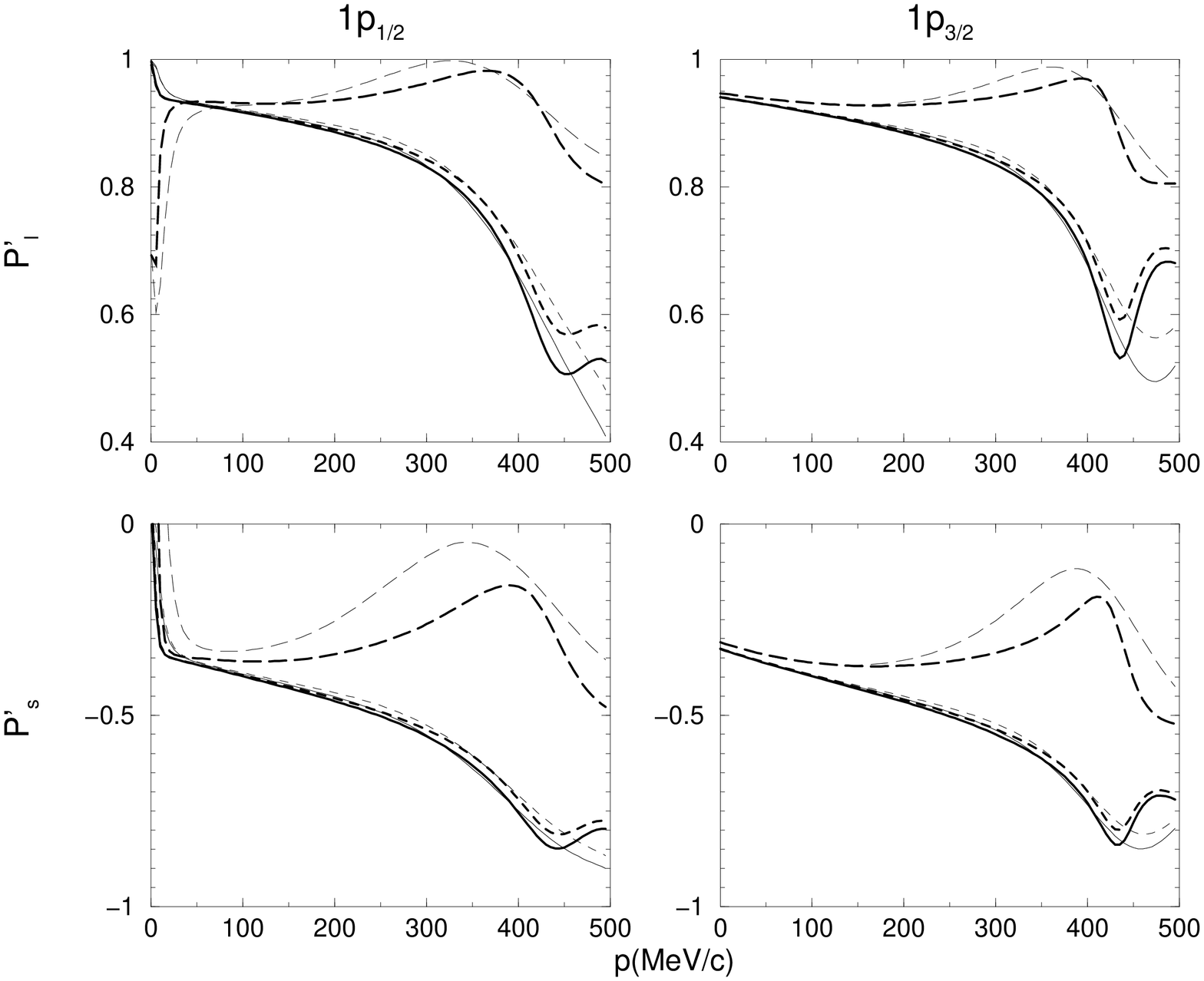}}} \par}
{\par\centering \resizebox*{0.8\textwidth}{0.4\textheight}{\rotatebox{270}
{\includegraphics{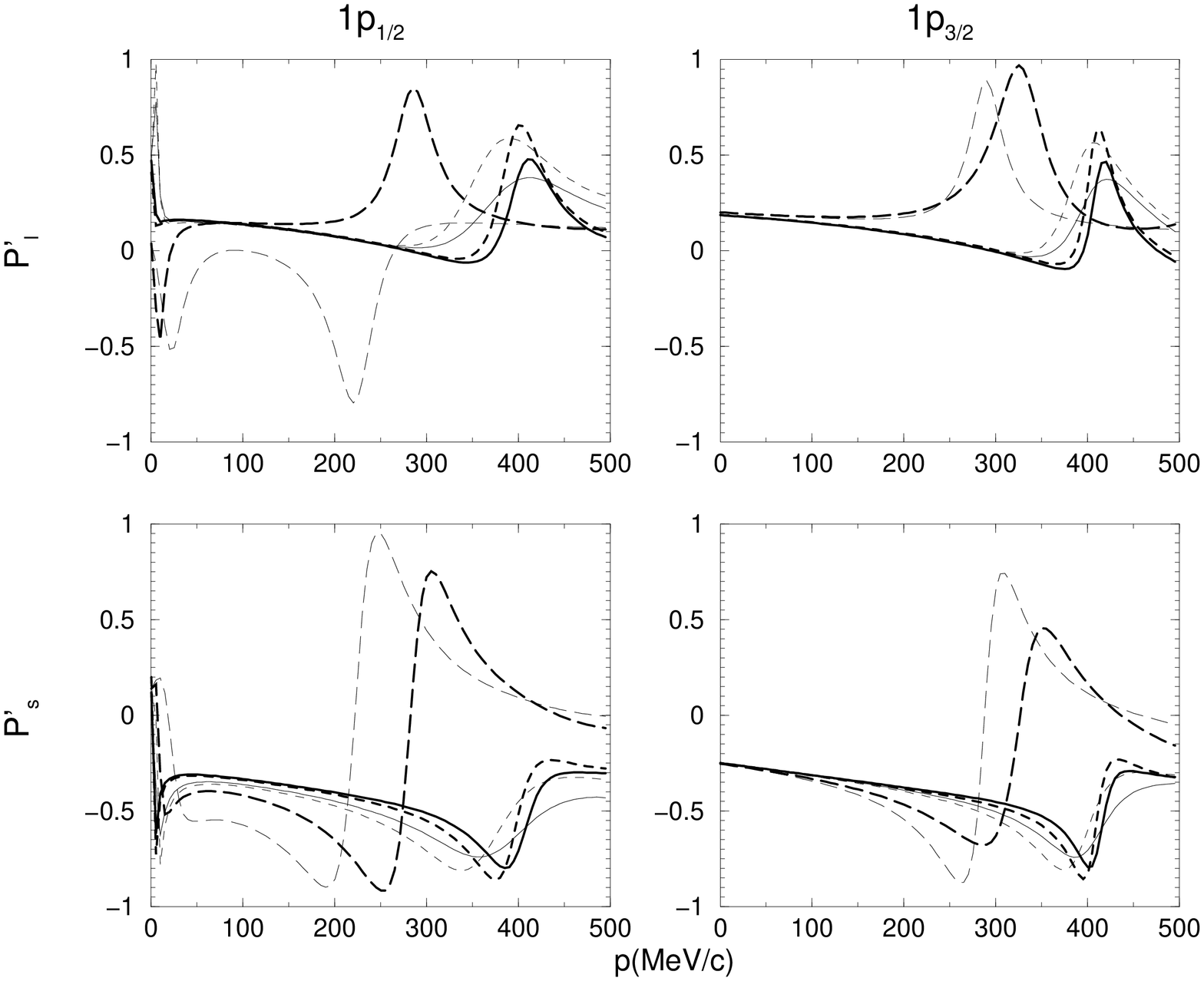}}} \par}
\caption{\label{sec33fig1}Transferred polarization asymmetries for longitudinal and sideways spin
directions. The results are for kinematics 1 (see text). Top panels correspond to 
\protect\( \theta_e =150^{o}\protect \)
and bottom panels to \protect\( \theta_e =30^{o}\protect \). right-hand panels correspond to the
\protect\( p_{1/2}\protect \) shell and left-hand panels to \protect\( p_{3/2}\protect \).
The labelling is the same as in Fig.~\ref{sec31fig1}.}
\end{figure}

\begin{figure}
{\par\centering \resizebox*{0.8\textwidth}{0.4\textheight}{\rotatebox{270}
{\includegraphics{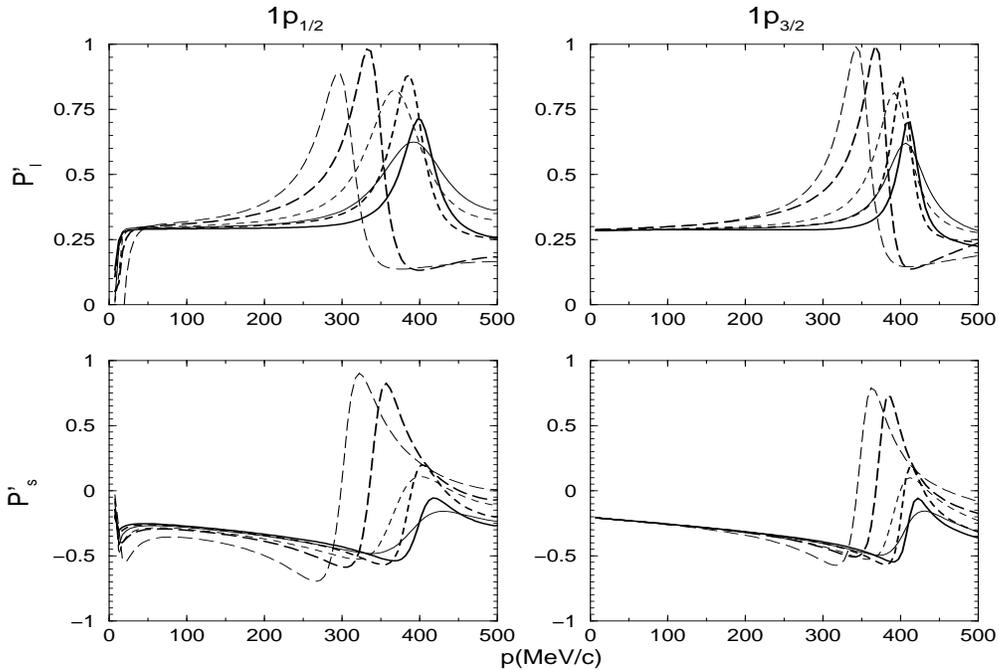}}} \par}
\caption{\label{sec33fig2}As for Fig.~\ref{sec33fig1}, but now for kinematics 2 and $\theta_e=23.4^0$}
\end{figure}

In Figs.~\ref{sec33fig1} and \ref{sec33fig2} we show the results for $P'_l$ and $P'_s$ corresponding
to kinematics 1 and 2, respectively. In both cases, $p_{1/2}$ (left-hand panels) and $p_{3/2}$
(right-hand panels) shells have been considered. Each graph presents six curves corresponding to the
six `off-shell' prescriptions discussed. Finally, in the case of kinematics 1 (Fig.~\ref{sec33fig1}),
results are shown for forward and backward electron scattering angles: 
$\theta_e=30^0$ ($\varepsilon=1$ GeV) (bottom panels)
and $\theta_e=150^0$ ($\varepsilon=324$ MeV) (top panels). 
For kinematics 2 (Fig.~\ref{sec33fig2}), only forward
electron scattering ($\theta_e=23.4^0$) ($\varepsilon=2.445$ GeV) has been considered.
The latter corresponds to experiment E89-033 performed at JLab~\cite{Mal00,E89003}.

From inspection of Fig.~\ref{sec33fig1} we may conclude the following:

\begin{itemize}
\item   
        The largest differences are produced when comparing the two Weyl gauge prescriptions, 
        $CC1^{(3)}$ and $CC2^{(3)}$, with the other four prescriptions corresponding to the Coulomb
        and Landau gauges. Note, in particular, the behaviour 
        of $CC1^{(3)}$ for $P'_l$ in the case of 
        the $p_{1/2}$ shell and forward electron scattering. On the other hand,
        the off-shell uncertainties are significantly reduced in the case of 
        the Coulomb and Landau gauges based prescriptions.
\item   
        Off-shell uncertainties are enhanced for forward electron scattering. Contrary to backward
        scattering where the transverse responses dominate, for forward electron scattering angles
        all of the kinematical factors are of similar order and consequently, off-shell effects are
        maximized.
\item   
        Within the backward scattering situation (top panels), the two polarization
        asymmetries are very similar for the Coulomb and Landau gauges up to roughly $p=300$ MeV/c. For
        higher $p$-values, the four prescriptions start to deviate, although their differences 
        are still relatively small.
\item
        In the case of forward scattering (bottom panels), we should distinguish between $P'_l$ and
        $P'_s$. In the former, Coulomb and Landau gauges with the two current choices give rise to
        almost identical results up to a missing momentum $p=300$ MeV/c. For $p>300$ MeV/c, the 
        results deviate from each other, their difference being considerably larger than the one observed
        for backward scattering. For $P'_s$ the differences amongst the four prescriptions associated
        with the Landau and Coulomb gauges start to be clearly visible at lower values of the 
        missing momentum $p$. 
\item   
        Finally, comparing the results for the $p_{1/2}$ and $p_{3/2}$ shells, we observe
        that the same qualitative behaviour is obtained for all of the off-shell prescriptions, with the
        exception of $CC1^{(3)}$ in $P'_l$. It is important to recall that the differences shown
        for both shells, which are larger for $\theta_e=30^0$, come from the dynamical relativistic effects.
        Within PWIA, due to factorization, the polarization asymmetries only depend on the
        single-nucleon responses, i.e., they are not affected by the bound wave function selected.  
        The different behaviour observed for both shells at very low $p$-values is directly connected
        to the quantum number $\overline{\ell}=0$ ($\overline{\ell}=2$) of the lower component in the
        $p_{1/2}$ ($p_{3/2}$) state (see~\cite{Cab98b} for details).

\end{itemize}

Transferred polarization asymmetries for kinematics 2 and $\theta_e=23.4^0$ are presented in 
Fig.~\ref{sec33fig2}. Some of the comments in the previous case can be also applied here; however,
some significant differences are also seen. In particular, note that, although the Weyl gauge 
still gives rise to
very different responses, the relative differences seen with the Coulomb and Landau gauges are not so
significant. This is connected with the fact that at very forward angles the 
off-shell uncertainty spread introduced by the Landau and Coulomb gauges is already very wide.
As in the previous case, off-shell ambiguities are maximized for $p_{1/2}$.

\begin{figure}
{\par\centering \resizebox*{0.8\textwidth}{0.4\textheight}{\rotatebox{270}
{\includegraphics{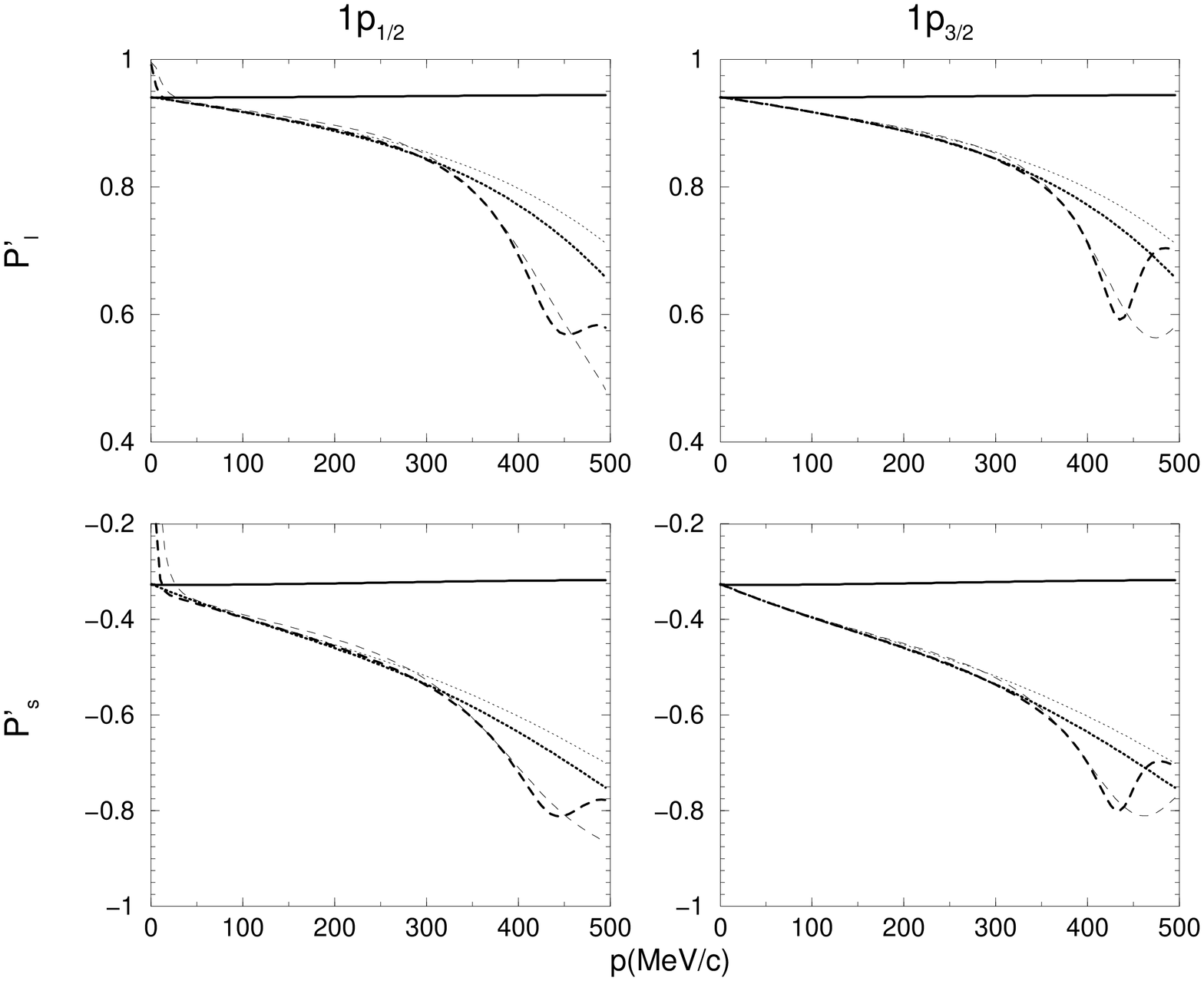}}} \par}
{\par\centering \resizebox*{0.8\textwidth}{0.4\textheight}{\rotatebox{270}
{\includegraphics{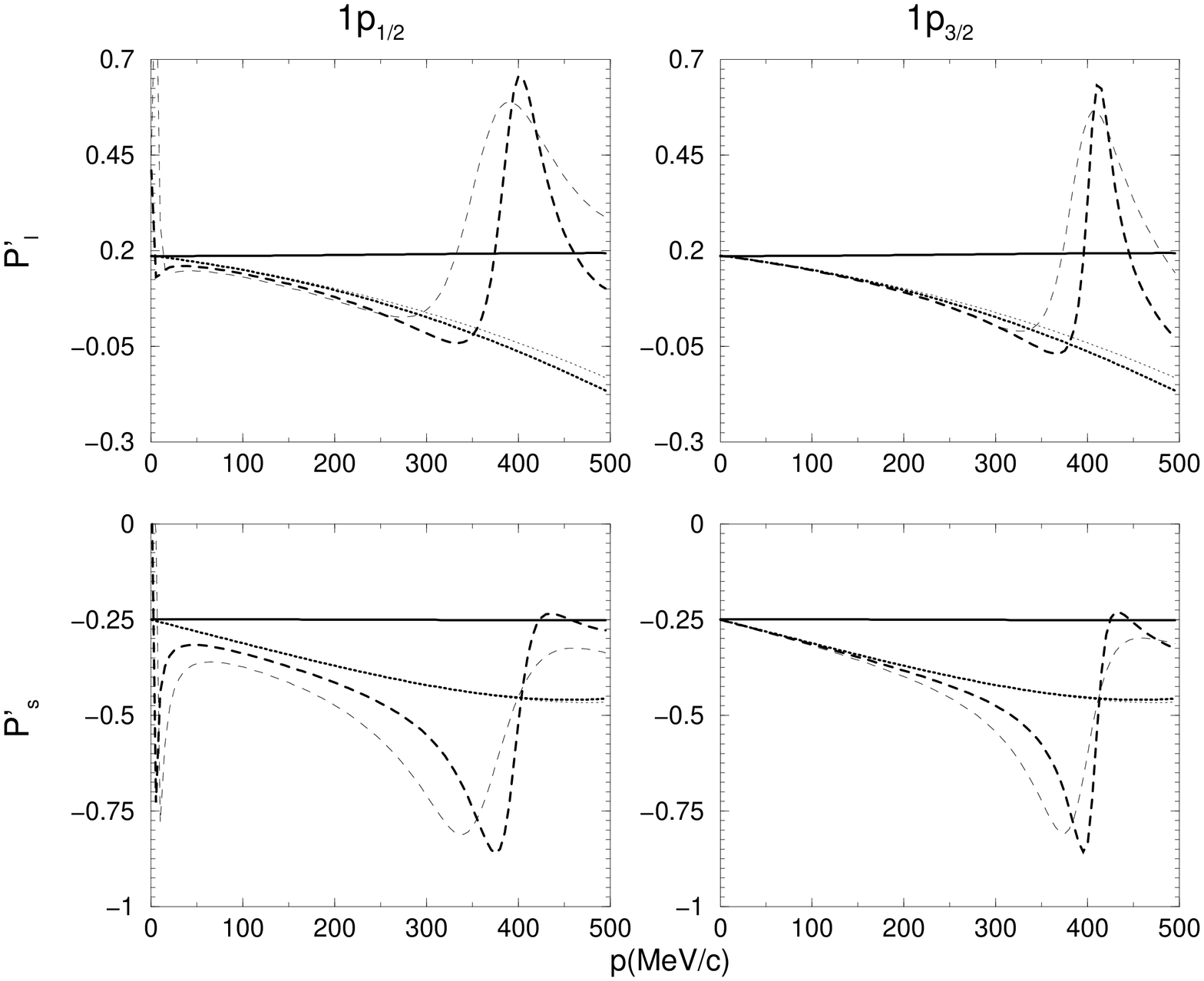}}} \par}
\caption{\label{sec33fig3} Transferred polarization asymmetries $P'_l$ and $P'_s$ for
kinematics 1. Fully-relativistic results corresponding to the Coulomb gauge (dashed lines)
are compared with their positive-energy projection contributions (dotted lines). Thin
lines correspond to the $CC1$ current operator and thick lines to $CC2$. 
We also show for comparison the static limit result (solid line).}
\end{figure}

The role played by the dynamical enhancement of the lower components in the bound nucleon wave
function is clearly seen in Fig.~\ref{sec33fig3}. Here 
the fully-relativistic RPWIA results (dashed lines) corresponding to the Coulomb gauge with
the $CC1$ and $CC2$ choices of the current, are compared with the transferred polarizations
obtained by projecting out the negative-energy components (dotted lines). As in previous figures,
results for $p_{1/2}$ (left-hand panels) and $p_{3/2}$ (right-hand panels) are shown. Results
correspond to kinematics 1, and forward ($\theta_e=30^0$, bottom panels) and backward 
($\theta_e=150^0$, top panels) angles have been considered. 

First, note the difference between the relativistic 
and projected results observed at very small missing momentum values for the $p_{1/2}$ shell.
As already mentioned (see~\cite{Cab98b} for details), this effect comes
directly from the quantum number $\overline{\ell}$ involved in the
lower component of the bound state wave function ($\overline{\ell}=0$ for $p_{1/2}$).
Moreover, it is also important to point out that
fully-relativistic and positive-energy projected
results typically do not differ appreciably, especially for backward angles,
for $p$-values up to $\sim 300$ MeV/c, although at forward angles some noticeable spread occurs at low $p$ for $P'_s$. For $p > 300$ MeV/c
relativistic and projected results start to deviate from each
other. This general behaviour is what one expects because of
the clear dominance of the positive-energy projection component of the momentum
distribution in the region $p\leq 300$ MeV/c~\cite{Cab98a,Cab98b}.
On the contrary, in the region of high missing momentum, $p>300$
MeV/c, the negative-energy components, $N_{uv}(p)$, $N_{vv}(p)$, are
similar to or even larger than $N_{uu}(p)$, and hence the effects of the
dynamical enhancement of the lower components in the bound
relativistic wave functions are clearly visible in the transfer 
polarization asymmetries.

From the results in
Fig.~\ref{sec33fig3} it is also clear that the dynamical effects are maximized in the
forward electron scattering situation. Here the differences between
fully-relativistic and projected results are significant even for
low/medium $p$-values, in particular for the sideways transfer
polarization, $P'_s$. As we know, the purely transverse responses dominate at backward angles, hence
the most relevant contributions to the polarization asymmetries
come from the transverse polarized responses $R^{T'}_l$ 
and/or $R^{T'}_s$ in the numerator, and 
from the unpolarized $R^T$ response in the denominator of Eq.~(\ref{eq39}).
From these three responses, only the small $R^{T'}_s$
is particularly sensitive to the effect of the negative-energy components
(see discussion in previous section).
In contrast, at forward angles
all of the kinematical factors that enter in the analysis of $(\vec{e},e'\vec{N})$ reactions
are of similar order, and hence the contribution of these responses 
that are more sensitive to dynamical relativistic effects 
is maximized. Accordingly, the role played by the 
negative-energy components of the bound relativistic wave function is emphasized
for transferred polarization asymmetries measured
at forward angles. Although not shown for simplicity,
these general comments also apply to the case of kinematics 2 and
$\theta_e=23.4^0$.

\begin{figure}
{\par\centering \resizebox*{0.8\textwidth}{0.6\textheight}{\rotatebox{270}
{\includegraphics{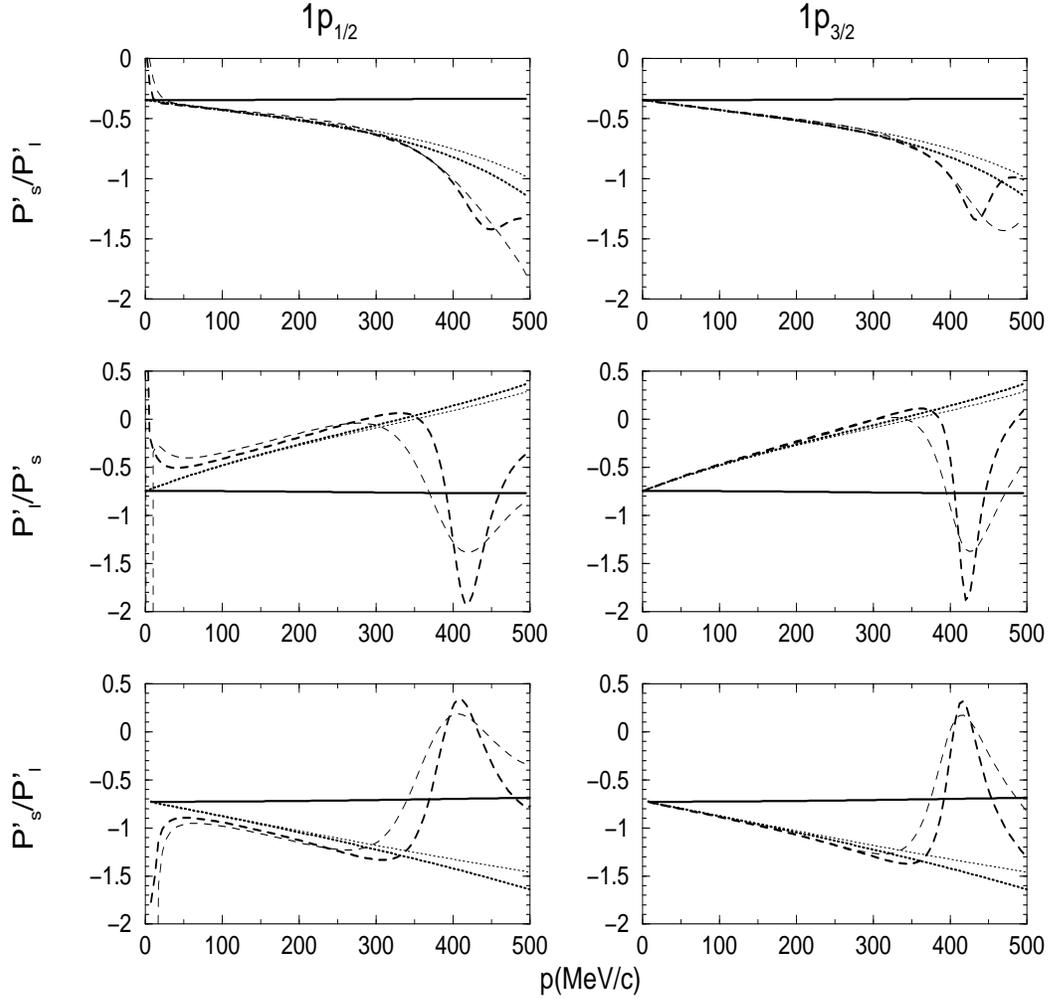}}} \par}
\caption{\label{sec33fig5}Ratios of polarization asymmetries for kinematics 1 and 2. Top (middle) panels
correspond to $\theta_e=150^0$ ($\theta_e=30^0$) and kinematics 1, while bottom panels correspond to $\theta_e=23.4^0$ and kinematics 2. Results for
the $p_{1/2}$ (left-hand panels) and $p_{3/2}$ (right-hand panels) shell are shown. The curves are labelled
as in Fig.~\ref{sec33fig3}. For comparison we also present the static limit result (solid line).}
\end{figure}

\begin{figure}
{\par\centering \resizebox*{0.8\textwidth}{0.6\textheight}{\rotatebox{270}
{\includegraphics{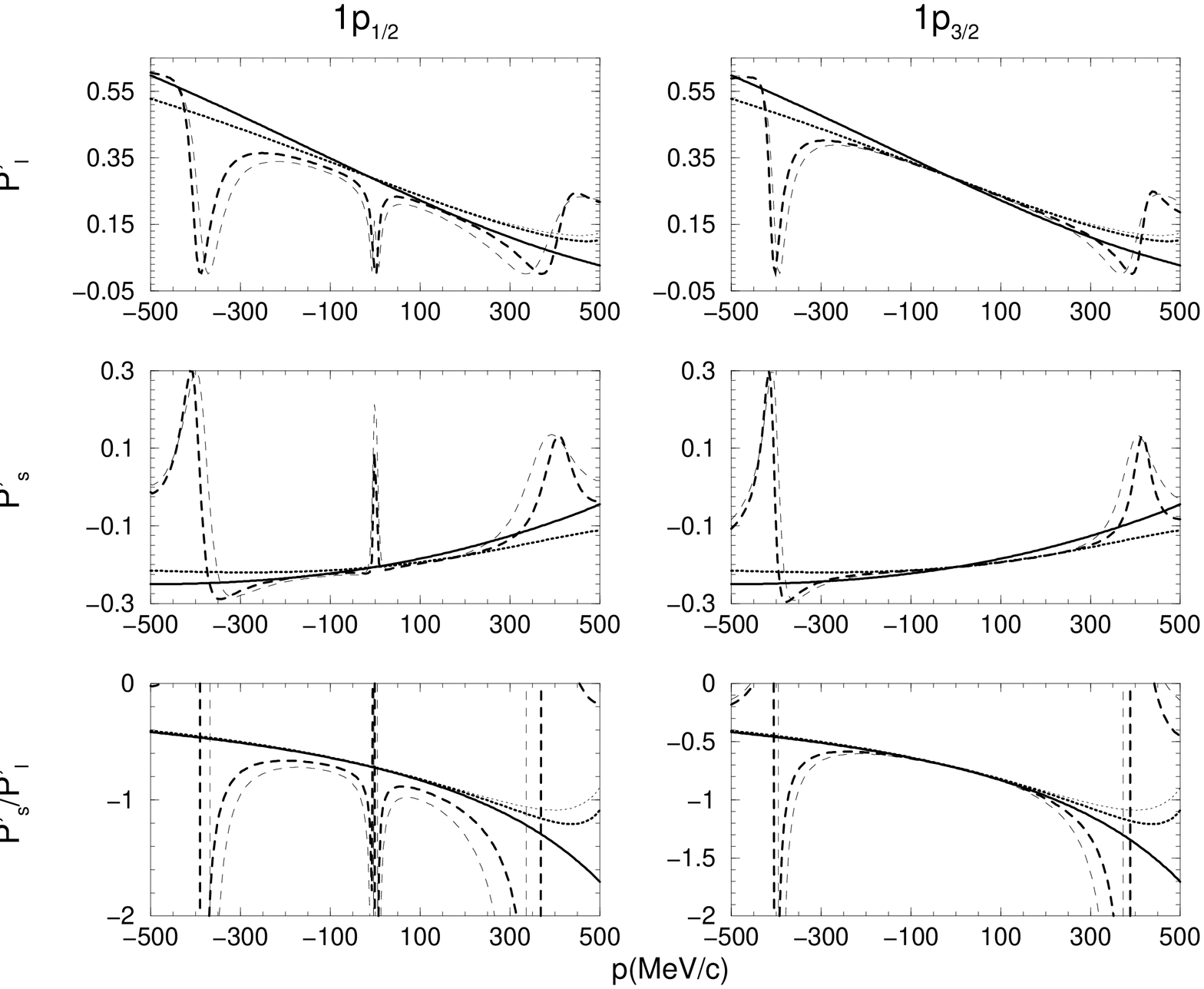}}} \par}
\caption{\label{sec33fig7}Transferred polarization asymmetries and ratio $P'_l/P'_s$ for kinematics 4
and $\varepsilon=2.445$ GeV. Results for
the $p_{1/2}$ (left-hand panels) and $p_{3/2}$ (right-hand panels) shell are shown. The curves are labelled
as in Fig.~\ref{sec33fig3} and Fig.~\ref{sec33fig5}.}
\end{figure}

Finally, as it has been proven to be a suitable observable for getting information on nucleon properties 
in the nuclear medium~\cite{Arn81}, we present in Fig.~\ref{sec33fig5} 
the ratio $P'_s/P'_l$ (or the reverse 
in the case that $P'_l$ goes through zero) for kinematics 1 and 2, respectively. As in previous figures,
we present fully-relativistic results ($CC1^{(0)}$ and $CC2^{(0)}$) and their positive-energy
projections, and we also show the results for the free-nucleon static limit 
as guidance. Following the general discussion already presented for the responses and polarization
asymmetries, we find:
\begin{itemize}
\item
        Relativistic dynamical effects show up mainly for $p>300$ MeV/c, modifying completely
        the slowly-varying behaviour of the positive-energy results. Nevertheless, there exist 
        already sizeable differences for low and moderate $p$-values for forward scattering angles.
\item
        Forward scattering angles enhance the sensitivity to dynamical relativistic effects, particularly
        for the $p_{1/2}$ shell, where relativistic effects show up clearly even at low/moderate values of 
        the missing momentum. For the $p_{3/2}$ shell these effects in
        the low and moderate $p$ region are much less important.
\item
        Relativistic and projected calculations deviate significantly from the free-nucleon static limit
        (solid line). At $p=100$ MeV/c the discrepancy is of the order of $\sim$20--50\% for kinematics
        1 (the largest values corresponding to forward angle) and $\sim$20--35\% for kinematics 2. 

\end{itemize}

To finish the discussion in this section, we show in Fig.~\ref{sec33fig7} the transferred polarization
asymmetries $P'_l$, $P'_s$ and ratio $P'_l/P'_s$, corresponding to parallel kinematics 4 and
beam energy $\varepsilon=2.445$ GeV (we do not present results for
kinematics 3 , since in that case the outgoing nucleon kinetic energy selected only allows us to reach low/medium
missing momentum values for which dynamical effects are expected to be
less relevant). We observe that, in concert with the
$(q,\omega)$-constant kinematics results, the highest differences
between projected and fully relativistic results appear in the high-$p$ 
region, although also for low and moderate $p$-values one can find
significant differences, mostly for the $P'_l$ observable. This is
connected with the fact that $R^{T'}_l$ shows more sensitivity to
dynamical relativistic effects than does $R^{TL'}_s$ in parallel
kinematics. Again, the differences for low and moderate $p$-values are
much reduced in the $p_{3/2}$ shell knockout case. Since the various
off-shell prescriptions lead to different results for the two transferred
polarization asymmetries and their ratio, clearly this impacts the
determination of the single-nucleon properties in the nuclear medium;
however, without yet having all of the relativistic effects in hand
--- specifically, since we lack a treatment of the relativistic FSI
problem within the same framework --- it is premature to state the
level of uncertainty that this entails. Upon completing the study
discussed in the Introduction~\cite{Cris3}, the impact of all of the
relativistic effects and the uncertainties they bring with them will
be quantified.


\section{Summary and conclusions}


In this paper we have studied the $A(\vec{e},e'\vec{N})B$ reaction within RPWIA. Our aim has
been to analyze dynamical relativistic effects associated with the bound nucleon wave
function for polarized hadronic
responses and transferred polarization asymmetries. Four different kinematical situations 
corresponding to $(q,\omega)$-constant (quasi-perpendicular) and parallel kinematics have 
been selected. In all cases results have been obtained for coplanar $\phi=0$ kinematics.
Polarized observables have been evaluated for various choices of the current
operator ($CC1$ versus $CC2$) and gauge (Coulomb, Landau and Weyl). As
prototypical cases proton knock-out from the $p_{1/2}$ and $p_{3/2}$
shells in $^{16}$O has provided the focus. Note that this implies that the
excitation energy $\cal{E}$ has been taken to be small, corresponding
to valence knock-out, and hence caution should be exercised in
extending the conclusions presented here to the case of studies at large missing energy.

From the results and discussion presented in the previous sections,
the following general conclusions can be drawn:

\begin{itemize}

\item
        The largest differences in the positive- and negative-energy projection components are
        produced by the prescriptions based on the Weyl gauge, particularly $CC1^{(3)}$. Moreover,
        off-shell uncertainties are much bigger for the negative-energy terms, whose relative
        contribution to the global responses is maximized by the $CC1$ choice of the current.
\item
        Within QEP kinematics, the responses $R^{T'}_s$ and
        particularly $R^{TL'}_l$ present the highest sensitivity to
        dynamical relativistic effects. Consequently, in light
        of the previous item, off-shell ambiguities are also bigger. On the contrary, the
        role of the negative-energy projections in the two remaining responses, $R^{T'}_l$ and 
        $R^{TL'}_s$, is almost negligible. Off-shell effects are also significantly reduced 
        except for the Weyl gauge in $R^{TL'}_s$. These results also connect with the general
        study performed in~\cite{Cris1} and also summarized above in Section~3.1, 
        where the responses $R^{T'}_l$ and $R^{TL'}_s$ 
        belong to class ``0'', while $R^{T'}_s$ and $R^{TL'}_l$ are
        class ``1'' responses. 
\item
        Dynamical relativistic effects and off-shell ambiguity in the
        two purely transverse $T'$-responses are shown to diminish for higher $q$-values
        within the QEP. In the case of the two interference $TL'$ responses and the prescriptions
        based on the Coulomb and Landau gauges, the effects introduced by the negative-energy 
        projections and off-shell uncertainties change slightly as $q$ varies within the QEP. 
        Particularly interesting is the case of $R^{TL'}_l$ and the $CC1^{(0)}$ prescription 
        for which dynamical relativistic effects do not depend on $q$.
\item
        Within parallel kinematics, the two surviving responses, $R^{T'}_l$ and $R^{TL'}_s$, 
        are particularly sensitive to dynamical relativistic effects --- much more than in the case
        of QEP kinematics. The spread of off-shell uncertainties is also wider in general.
        These results are directly connected with the fact that in parallel kinematics varying
        the missing momentum $p$ means moving far away from the QEP, and consequently the role
        of negative-energy projections and off-shell effects are enhanced.
\item   
        Dynamical relativistic effects evaluated at $p=\pm 100$ MeV/c (which 
        corresponds approximately to the peaks of the responses), are shown to be
        larger for positive $p$-values than for negative $p$-values. This is so because positive $p$-values
        correspond to being further away from the QEP than the
        negative $p$-values with the same $|p|$. This seems to be correlated with the value of $|y|/q$. In fact, this also explains why the role of negative-energy
        projections, evaluated at the peaks of
        the responses, diminishes for larger outgoing nucleon momentum (kinematics~4), 
        and why the behaviour shown for positive and negative $p$-values tends to 
        be similar for increasing $p_N$.
\item
        The influence of dynamical relativity and off-shell effects in the transferred polarization
        asymmetries is clearly maximized in the case of forward electron scattering.
        At high $p$-values the dynamical enhancement of the lower components in
        the bound nucleon wave function may modify completely the structure of the
        polarization asymmetries. In contrast, for $p<300$ MeV/c the negative-energy projections
        are in general well under control, and do not introduce significant effects, particularly
        for backward electron scattering. 
\item   
        The dynamical relativistic results obtained for the 
        ratio between longitudinal and sideways transferred polarizations 
        deviate significantly from the static limit approximation when using $(q,\omega)$-constant kinematics, even at low/medium $p$-values. Within parallel kinematics the discrepancies with 
        the static limit diminish, especially in the case of the $CC2$ current. 
        However, even in this case, one should be very cautious before precise 
        information on the in-medium nucleon form factors
        can be revealed from the analysis of $(\vec{e},e'\vec{N})$ reactions.
\item   
        Dynamical relativistic effects and off-shell uncertainties in the response 
        functions are significantly reduced for a proton being
        knocked-out from the $p_{3/2}$ shell
        which, having lower components of $d$-wave character rather
        than of $s$-wave character as in the $p_{1/2}$ shell case, is natural. 
        The transferred polarization asymmetries evaluated for both shells, $p_{1/2}$ and
        $p_{3/2}$, in general display a similar structure, at high $p$ showing an oscillatory
        behaviour that within RPWIA comes solely from the role played by the negative-energy components.
        This implies a crucial difference with respect to the standard PWIA and non-relativistic analyses.

\end{itemize}

Finally, although being aware of the significant modifications that FSI 
may introduce in the analysis, we are rather confident that
the high sensitivity of polarization-related observables 
to negative-energy projections shown within RPWIA will probably also be maintained within more elaborated relativistic
distorted-wave impulse approximation (RDWIA) calculations.
Work along this line is presently in progress.

\subsection*{Acknowledgements}
This work was partially supported by funds provided by DGICYT (Spain) 
under Contracts Nos. PB/98-1111, PB/98-0676 and the Junta
de Andaluc\'{\i}a (Spain) and in part by the U.S. Department of Energy under
Cooperative Research Agreement No. DE-FC02-94ER40818.
M.C.M acknowledges support from a fellowship from the Fundaci\'on
C\'amara (University of Sevilla). J.A.C. also acknowledges financial support from MEC (Spain) for a sabbatical stay at MIT (PR2001-0185). The authors thank E. Moya, J.M. Ud\'{\i}as and J.R. Vignote for
their helpful comments.


\appendix


\section*{Appendix A}


In this appendix we present the formalism needed to evaluate the positive- and
negative-energy projection contributions to the hadronic tensor in the case of recoil nucleon
polarization measurements.

The nucleon current matrix element in RPWIA is given by
\be
\langle J^\mu \rangle \equiv \overline{u}(\np_N,s_N)\hat{J}^\mu \Psi_\kappa^m(\np)
\label{ap1}
\ee
with $\Psi_\kappa^m(\np)$ a relativistic bound nucleon wave function in momentum space. As
already mentioned, this wave function has coupling to the negative-energy Dirac spinors $v$.
Making use of the completeness relation for Dirac spinors or introducing positive- and
negative-energy projector operators in Eq.~(\ref{ap1}), the nucleon current matrix element
can be decomposed into two terms, $\langle J^\mu\rangle=\langle J^\mu\rangle_u-\langle J^\mu\rangle_v$,
the former (latter) coming from the positive- (negative-) energy projector involving the Dirac 
spinors $u(\np, s)$ ($v(\np, s)$). These terms can be written in the form
\ba
\langle J^\mu\rangle_u&=&\sqrt{\frac{\overline{E}+M_N}{2M_N}}(-i)^\ell\alpha_\kappa (p)
        \sum_s\left[\overline{u}(\np_N,s_N)\hat{J}^\mu u(\np,s)\right]
        \langle s|\Phi_\kappa^m(\hat{\np})\rangle
\label{ap2}\\
\langle J^\mu\rangle_v&=&-\sqrt{\frac{\overline{E}+M_N}{2M_N}}(-i)^\ell\beta_\kappa (p)
        \sum_s\left[\overline{u}(\np_N,s_N)\hat{J}^\mu v(\np,s)\right]
        \langle s|\Phi_{-\kappa}^m(\hat{\np})\rangle\, ,
\label{ap3}
\ea
where use has been made of the explicit expression of the relativistic bound nucleon wave function 
in momentum space as given in~\cite{Cab98a}. The term 
$\langle s|\Phi_{\pm\kappa}^m(\hat{\np})\rangle$
indicates spin projection of the bispinors $\Phi_{\pm\kappa}^m$ on a spin state $|\frac{1}{2} s\rangle$, 
and the radial functions in momentum space $\alpha_\kappa (p)$ and $\beta_\kappa (p)$ 
were introduced in~\cite{Cab98a} and their explicit expressions are
given in Eqs.~(\ref{eq21},\ref{eq22}).

Now we can proceed to evaluate the three contributions that enter in the hadronic tensor $W^{\mu\nu}$ in Eq.~(\ref{eq14}).
\begin{itemize}
\item {\bf Positive-energy projection tensor (\( W_{P}^{\mu \nu } \))}.
\be
W^{\mu \nu }_{P} = \frac{2}{2j+1}\sum _{m}\langle J^{\mu }\rangle _{u}^{\ast}
\langle J^{\nu }\rangle _{u}\, .
\label{ap4}
\ee
Introducing the expression (\ref{ap2}) one simply gets
\be
W_{P}^{\mu\nu}=\frac{2}{2j+1}\frac{\overline{E}+M_{N}}{2M_{N}}\mid \alpha _{\kappa}(p)\mid ^{2}
\sum _{ss'}{\cal W}_{ss'}^{\mu\nu}
\sum _{m}\langle s\mid \Phi ^{m}_{\kappa}\rangle ^{*}\langle s'\mid \Phi _{\kappa}^{m}\rangle \, ,
\label{ap5}
\ee
where the single-nucleon tensor \( {\cal W}^{\mu \nu }_{ss'} \) is defined as
\be
{\cal W}_{ss'}^{\mu\nu}=\left[\overline{u}(\np_{N},s_{N})\hat{J}^{\mu }
u(\np,s)\right]^{\ast}\left[\overline{u}(\np_{N},s_{N})\hat{J}^{\nu }u(\np,s')\right] \, .
\label{ap6}
\ee
Finally, using the relation
\begin{equation}
\label{eq:sumaenm}
\sum _{m}\langle s\mid \Phi ^{m}_{\kappa}\rangle ^{\ast}
\langle s'\mid \Phi _{\kappa}^{m}\rangle =\frac{2j+1}{8\pi }\delta _{ss'}\, ,
\end{equation}
the hadronic tensor contribution $W^{\mu\nu}_P$ can be written in the form shown in
Eq.~(\ref{eq15}).

\item {\bf Negative-energy projection tensor (\( W^{\mu \nu }_{N} \)).}\\
Analogously to the previous case we have
\ba
W^{\mu \nu }_{N} & = & \frac{2}{2j+1}\sum _{m}\langle J^{\mu }
\rangle _{v}^{*}\langle J^{\nu }\rangle _{v}\\
& = &\frac{2}{2j+1}\frac{\overline{E}+M_{N}}{2M_{N}}
\mid \beta_{\kappa}(p)\mid ^{2}\sum _{ss'}{\cal Z}_{ss'}^{\mu \nu }
\sum _{m}\langle s\mid \Phi ^{m}_{-\kappa}\rangle ^{\ast}\langle s'\mid \Phi _{-\kappa}^{m}\rangle \, ,
\label{ap7}
\ea
with the single-tensor \( {\cal Z}_{ss'}^{\mu \nu } \) defined as
\be
{\cal Z}_{ss'}^{\mu \nu }=
\left[\overline{u}(\np_{N},s_{N})\hat{J}^{\mu}v(\np,s)\right]^{\ast}
\left[\overline{u}(\np_{N},s_{N})\hat{J}^{\nu}v(\np,s')\right] \, .
\label{ap8}
\ee
Using again the relation (\ref{eq:sumaenm}), which means that only the spin-diagonal
contributions survive, the hadronic tensor component $W^{\mu\nu}_N$ factorizes into two terms
as given in Eq.~(\ref{eq16}).

\item {\bf Crossed tensor (\( W^{\mu \nu }_{C} \))}. 
\be
W^{\mu \nu }_{C}= -\frac{2}{2j+1}\sum _{m}\left[\langle J^{\mu }
\rangle _{u}^{*}\langle J^{\nu }\rangle _{v}+\langle J^{\mu }\rangle _{v}^{*}
\langle J^{\nu }\rangle _{u}\right] \, .
\label{ap9}
\ee
Using the results shown in Eqs.~(\ref{ap2},\ref{ap3})
and taking into account that
the radial functions \( \alpha_{\kappa}(p) \) and \( \beta_{\kappa}(p) \) in Eqs.~(\ref{eq21},\ref{eq22})
are both real, the hadronic crossed tensor can be written as 
\be
W^{\mu \nu }_{C}=\frac{\overline{E}+M_{N}}{8\pi M_{N}}\alpha _{\kappa}(p)\beta_{\kappa}(p)
\sum_{ss'}\langle s'\mid \frac{\nsigma\cdot\np}{p}\mid s\rangle \left[{\cal I}^{\mu\nu}_{s's}
+\left({\cal I}_{ss'}^{\nu \mu }\right)^{\ast}\right]\, ,
\label{ap10}
\ee
where the single-nucleon tensor \( {\cal I}_{s's}^{\mu\nu} \) is given by
\be
{\cal I}_{ss'}^{\mu\nu}=\left[\overline{u}(\np_{N},s_{N})\hat{J}^{\mu}
u(\np ,s)\right]^{\ast}\left[\overline{u}(\np_{N} ,s_{N})\hat{J}^{\nu}v(\np ,s')\right] \, ,
\label{ap11}
\ee
and the sum over \( m \) has been performed using the relation
\be
\sum _{m}\langle s\mid \Phi ^{m}_{\kappa}\rangle ^{\ast}\langle s'\mid \Phi _{-\kappa}^{m}\rangle 
=-\frac{2j+1}{8\pi}\langle s'\mid \frac{\nsigma\cdot\np}{p}\mid s\rangle \, .
\label{ap12}
\ee

The single-nucleon tensors \( {\cal I}^{\mu \nu }_{s's} \), \( ({\cal I}^{\nu \mu }_{ss'})^{*} \) can
be written using traces in the general form
\be
{\cal I}_{s's}^{\mu \nu }=\frac{1}{16M_{N}^{2}}
Tr\left[ \gamma _{5}(\delta _{ss'}+\gamma _{5}\cinvslash )(\pslash +M_{N})
\overline{J}^{\mu }(\pnslash +M_{N})(1+\gamma _{5}\snslash )J^{\nu }\right] \, ,
\label{ap13}
\ee
where we have introduced the pseudovector
$C_{ss'}^{\mu }=\langle \np,s'\mid \gamma ^{\mu }\gamma ^{5}\mid \np,s\rangle$, which
reduces to the four-spin of the bound nucleon \( S^{\mu }_{L} \) in the diagonal case $s=s'$.

Splitting the tensor ${\cal I}_{s's}^{\mu\nu}$ into its symmetric (${\cal S}$)
and antisymmetric (${\cal A}$) terms this can be written in general
\[
{\cal I}_{s's}^{\mu \nu }={\cal S}^{\mu \nu }_{s's}+{\cal A}_{s's}^{\mu \nu }(S_{N})+
\delta _{ss'}\left[{\cal A}^{\mu \nu }+{\cal S}^{\mu \nu }(S_{N})\right]\, .
\]
This means that the dependence on the recoil nucleon polarization is entirely contained in
the antisymmetric spin non-diagonal and symmetric spin-diagonal terms. Carrying out the
sum over \( s \) and \( s' \) indicated in Eq.~(\ref{ap10}), and taking into account the 
relation
\be
\langle s\mid \nsigma\cdot\np\mid s\rangle =
-\langle -s\mid \nsigma\cdot\np\mid -s\rangle \, ,
\label{ap14}
\ee
it can easily be shown that the spin diagonal terms give no contribution.
Hence the whole dependence on the outgoing nucleon four-spin $S_N^\mu$ is
contained in the surviving antisymmetric part of the tensor, ${\cal A}^{\mu\nu}_{s's}$.
A similar analysis can be also applied to \( \left({\cal I}_{ss'}^{\nu\mu}\right)^{\ast} \).
Thus, we can write
\be
{\cal I}^{\mu\nu}_{s's}+\left({\cal I}_{ss'}^{\nu \mu }\right)^{\ast}\equiv
{\cal R}^{\mu\nu}_{s's}=\left({\cal R}_{s's}^{\mu\nu}\right)^{sym}+
        \left({\cal R}_{s's}^{\mu\nu}\right)^{ant}
\label{ap15}
\ee
with
\ba
\left({\cal R}_{s's}^{\mu\nu}\right)^{sym}&=&\frac{1}{8M_{N}}Tr\left[ 
\cinvslash \overline{J}^{\mu }(\pnslash +M_{N})J^{\nu }\right] \label{ap16} \\
\left({\cal R}_{s's}^{\mu\nu}\right)^{ant}&=&\frac{1}{8M_{N}}Tr\left[ 
\cinvslash \overline{J}^{\mu }\gamma _{5}\snslash (\pnslash +M_{N})J^{\nu }\right] \, .
\label{ap17}
\ea
The sum over \( s,s' \) can be explicitly perfomed, yielding
\be
\label{eq:sumrmunussp}
\sum _{ss'}\langle s'\mid \frac{\nsigma\cdot\np}{p}
\mid s\rangle {\cal R}^{\mu \nu }_{s's}=
({\cal R}^{\mu \nu }_{++}-{\cal R}_{--}^{\mu \nu })\cos\theta 
+({\cal R}_{+-}^{\mu \nu }e^{-i\phi }+{\cal R}_{-+}^{\mu \nu }e^{i\phi })\sin\theta \, ,
\ee
with \( \theta ,\phi \) the angles defining the direction of the bound nucleon
momentum $\np$.

Finally, using the spin precession technique developed in~\cite{Cab93}, the different
components of the spin-dependent tensor ${\cal R}^{\mu\nu}_{s's}$ can be written in terms of 
a general spin-diagonal tensor ${\cal R}^{\mu\nu}(S_L)$ in Eq.~(\ref{eq27}) whose spin 
four-vector $S_L^\mu$ is quantized
with respect to a generic direction defined by the angles $(\theta_L,\phi_L)$ in the
laboratory frame or $(\theta_R,\phi_R)$ in the frame where the nucleon is at rest (see \cite{Cab93} for details). The final expression for the tensor ${\cal N}^{\mu\nu}$ 
introduced in Eq.~(\ref{eq17}) is 
as given in Eq.~(\ref{eq26}). The components in the laboratory frame of the four-vector $S^\mu_L$ 
needed to evaluate ${\cal N}^{\mu\nu}$ are summarized in Table~C.1 of~\cite{Cab98a}.

\end{itemize}

\underline{\bf Momentum distribution components}\\

The explicit expressions of the momentum distribution components
$N_{uu}(p)$, $N_{vv}(P)$ and $N_{uv}(P)$ introduced in Eqs.~(\ref{eq15}-\ref{eq17}) are
\ba
N_{uu}(p)&=&\frac{\overline{E}+M_{N}}{8\pi M_{N}}\mid \alpha _{\kappa}(p)\mid ^{2} \label{eq18}\\
N_{vv}(p)&=&\frac{\overline{E}+M_{N}}{8\pi M_{N}}\mid \beta _{\kappa}(p)\mid ^{2} \label{eq19} \\
N_{uv}(p)&=&-\frac{\overline{E}+M_{N}}{4\pi M_{N}}\alpha _{\kappa}(p)\beta_{\kappa}(p) \label{eq20}\, ,
\ea
with the functions $\alpha_\kappa$ and $\beta_\kappa$ given by
\ba
\alpha_\kappa (p) &=& g_\kappa (p) -\frac{p}{\overline{E}+M_N}S_\kappa f_\kappa (p) \label{eq21}\\
\beta_\kappa (p) &=& \frac{p}{\overline{E}+M_N}g_\kappa (p)-S_\kappa f_\kappa (p) \label{eq22}\, ,
\ea
and $g_\kappa$, $f_\kappa$ the Bessel transforms of the standard upper and lower radial
functions of the bound nucleon wave function.


\section*{Appendix B}


In this appendix we present explicit expressions for the single-nucleon tensors
${\cal W}^{\mu\nu}$, ${\cal Z}^{\mu\nu}$ and ${\cal R}^{\mu\nu}$ that enter in the analysis
of \( A(\vec{e},e'\vec{N}) \) reactions within RPWIA. We show the results corresponding
to the two current operators defined as $CC1$ and $CC2$ (see~\cite{Cris1,Cab93,For83}). 
For simplicity we only
show the expressions for the antisymmetric parts of the tensors which contain the entire
dependence on the recoil nucleon polarization. The symmetric terms, giving rise to the
unpolarized responses, are half the ones given in~\cite{Cab98a}.

\begin{itemize}

\item {\bf $CC1$ current} 

\ba
{\cal W}_{A}^{\mu \nu } & = & \frac{i}{2M^{2}_{N}}\left\{
        M_{N}(F_{1}+F_{2})^{2}\varepsilon ^{\alpha\beta\mu\nu}S_{N_{\alpha }}\overline{Q}_{\beta }
                                \right. \nonumber \\
 & + &\left. (F_{1}+F_{2})\frac{F_{2}}{2M_{N}}\left[ 
        (\overline{P}+P_{N})^{\mu }\varepsilon ^{\alpha \beta \gamma \nu }-
        (\overline{P}+P_{N})^{\nu }\varepsilon ^{\alpha \beta \gamma \mu }\right] 
        \overline{P}_{\alpha }P_{N_{\beta }}S_{N_{\gamma }}\right\}
\label{ap18}
\ea

\ba
{\cal Z}_{A}^{\mu \nu } & = & \frac{i}{2M^{2}_{N}}\left\{
        M_{N}(F_{1}+F_{2})^{2}\varepsilon ^{\alpha \beta \mu \nu }
        (P_{N}+\overline{P})_{\alpha }S_{N_{\beta }}\right. \nonumber \\
 & + &\left. (F_{1}+F_{2})\frac{F_{2}}{2M_{N}}\left[ (\overline{P}+P_{N})^{\mu }
        \varepsilon ^{\alpha \beta \gamma \nu }-(\overline{P}+P_{N})^{\nu }
        \varepsilon ^{\alpha \beta \gamma \mu }\right] \overline{P}_{\alpha }
        P_{N_{\beta }}S_{N_{\gamma }}\right\}
\label{ap19}
\ea

\ba
{\cal R}_{A}^{\mu \nu }(S_{L}) & = & \frac{i}{2M_{N}}\left\{
        M_{N}(F_{1}+F_{2})^{2}\varepsilon ^{\alpha \beta \mu \nu }S_{L_{\alpha }}S_{N_{\beta }}\right.
        \nonumber \\
 & + &\left. (F_{1}+F_{2})\frac{F_{2}}{2M_{N}}\left[ (\overline{P}+P_{N})^{\mu }
        \varepsilon ^{\alpha \beta \gamma \nu }-(\overline{P}+P_{N})^{\nu }
        \varepsilon ^{\alpha \beta \gamma \mu }\right] S_{L_{\alpha }}
        P_{N_{\beta }}S_{N_{\gamma }}\right\} \nonumber \\
& & 
\label{ap20}
\ea

\item {\bf $CC2$ current}

\ba
{\cal W}_{A}^{\mu \nu } & = & \frac{i}{2M^{2}_{N}}\left\{
        M_{N}F_{1}^{2}\varepsilon ^{\alpha \beta \mu \nu }S_{N_{\alpha }}\overline{Q}_{\beta }\right.
\nonumber \\
 & + &\left. \frac{F_{1}F_{2}}{2M_{N}}\left[ (\right. Q^{\mu }
        \varepsilon ^{\alpha \beta \gamma \nu }-Q^{\nu }\varepsilon ^{\alpha \beta \gamma \mu })
        S_{N_{\alpha }}P_{N_{\beta }}\overline{P}_{\gamma } \right.
\nonumber \\
& + &\left. 
2\varepsilon ^{\alpha \beta \mu \nu }(\overline{P}\cdot QP_{N_{\alpha }}
        S_{N_{\beta }}-M^{2}_{N}Q_{\alpha }S_{N_{\beta }}\left. )\right] \right. 
\nonumber \\
 & - &\left. \frac{F^{2}_{2}}{4M_{N}}\left[ (\right. Q^{\mu }
        \varepsilon ^{\alpha \beta \gamma \nu }-Q^{\nu }
        \varepsilon ^{\alpha \beta \gamma \mu })S_{N_{\alpha }}(\overline{P}+P_{N})_{\beta }
        Q_{\gamma } \right. 
\nonumber \\
 &+  & \left. 
\varepsilon ^{\alpha \beta \mu \nu }(2\overline{P}\cdot Q S_{N_{\alpha }}
        Q_{\beta }+Q^{2}(\overline{P}+P_{N})_{\alpha }S_{N_{\beta }}\left. )\right] \right\}
\label{ap21}
\ea

\ba
{\cal Z}_{A}^{\mu \nu } & = & \frac{i}{2M^{2}_{N}}\left\{
        M_{N}F_{1}^{2}\varepsilon ^{\alpha \beta \mu \nu }
        (\overline{P}+P_{N})_{\alpha }S_{N_{\beta }}\right. 
\nonumber \\
 &+  &\left. \frac{F_{1}F_{2}}{2M_{N}}\left[ (\right. Q^{\mu }
        \varepsilon ^{\alpha \beta \gamma \nu }-Q^{\nu }
        \varepsilon ^{\alpha \beta \gamma \mu })S_{N_{\alpha }}P_{N_{\beta }}
        \overline{P}_{\gamma }\right. 
\nonumber \\
 & + &\left.
 2\varepsilon^{\alpha\beta\mu\nu}(\overline{P}\cdot QP_{N_{\alpha }}
        S_{N_{\beta }}-M^{2}_{N}S_{N_{\alpha }}Q_{\beta }\left. )\right] \right. 
\nonumber \\
 & + &\left. \frac{F^{2}_{2}}{4M_{N}}\left[ (\right. Q^{\mu }
        \varepsilon ^{\alpha \beta \gamma \nu }-Q^{\nu }
        \varepsilon ^{\alpha \beta \gamma \mu })S_{N_{\alpha }}
        \overline{Q}_{\beta }Q_{\gamma } \right. 
\nonumber \\
 & + &\left. 
\varepsilon ^{\alpha \beta \mu \nu }(2\overline{P}\cdot QQ_{\alpha }
        S_{N_{\beta }}+Q^{2}\overline{Q}_{\alpha }S_{N_{\beta }}\left. )\right] \right\}
\label{ap22}
\ea

\ba
{\cal R}_{A}^{\mu \nu }(S_{L}) & = & \frac{i}{2M_{N}}\left\{
        M_{N}F_{1}^{2}\varepsilon ^{\alpha \beta \mu \nu }S_{L_{\alpha }}S_{N_{\beta }}\right. 
\nonumber \\
 & + &\left. \frac{F_{1}F_{2}}{2M_{N}}\left[ (\right. P_{N}^{\mu }
        \varepsilon ^{\alpha \beta \gamma \nu }-P_{N}^{\nu }
        \varepsilon ^{\alpha \beta \gamma \mu })S_{L_{\alpha }}
        S_{N_{\beta }}Q_{\gamma }\right. 
\nonumber \\
 &+  &\left. \varepsilon ^{\alpha \beta \mu \nu }(S_{L}\cdot QP_{N_{\alpha }}
        S_{N_{\beta }}-P_{N}\cdot S_{L}S_{N_{\alpha }}
        Q_{\beta }-Q\cdot S_{N}S_{L_{\alpha }}P_{N_{\beta }}\left. )\right] \right. 
\nonumber \\
 &-  &\left. \frac{F^{2}_{2}}{4M_{N}}\left[ (\right. Q^{\mu }
        \varepsilon ^{\alpha \beta \gamma \nu }-Q^{\nu }
        \varepsilon ^{\alpha \beta \gamma \mu })S_{L_{\alpha }}S_{N_{\alpha }}
        Q_{\gamma }\right. 
\nonumber \\
 &+ & \left. \varepsilon ^{\alpha \beta \mu \nu }(2Q\cdot S_{N}S_{L_{\alpha }}
        Q_{\beta }-Q^{2}S_{L_{\alpha }}S_{N_{\beta }}\left. )\right] \right\}
\label{ap23}
\ea

\end{itemize}

Finally, in Table~\ref{tab2} we show the components of the recoil nucleon spin 4-vector $S_N^\mu$ 
in the laboratory
frame along the three directions defined by the axes $\nl$, $\ns$ and $\nn$.

\begin{table}
{\centering \begin{tabular}{ccccc}
\hline
Axis & \( \mu=0 \) & \( \mu=1 \)& \( \mu=2 \) & \( \mu=3 \)
\\
\hline
\\
$\nl$ & $\sqrt{\gamma_N^2-1}$ & $\gamma_N\sin\theta_N$ & 0 & $\gamma_N\cos\theta_N$ 
\\
$\ns$ & 0 & $\cos\theta_N$ & 0 & $-\sin\theta_N$ 
\\
$\nn$ & 0 & 0 & 1 & 0 \\
\hline 
\end{tabular}\par}

\caption{\label{tab2} Components of the spin 4-vector $S_N^\mu(\nl,\ns,\nn)$. The notation 
$\gamma_N\equiv E_N/M_N$ has been introduced.}
\end{table}



\end{document}